\documentclass[lettersize,journal]{IEEEtran}
\usepackage{amsmath,amsfonts}
\usepackage{algorithmic}
\usepackage{algorithm}
\usepackage{array}
\usepackage[caption=false,font=normalsize,labelfont=sf,textfont=sf]{subfig}
\usepackage{textcomp}
\usepackage{stfloats}
\usepackage{url}
\usepackage{verbatim}
\usepackage{amssymb}
\usepackage{graphicx}
\usepackage{cite}
\begin{document}

\title{Range Ambiguity Identification and Elimination by Leveraging Temporal Broadening Effect for THz Sensing}

\author{Saira Rafique, Ahmed Naeem, Chong Han~\IEEEmembership{Senior~Member,~IEEE}, Hüseyin~Arslan~\IEEEmembership{Fellow,~IEEE}
\thanks{Saira Rafique is with the IPR and License Agreements Department, Vestel Electronics, 45030
Manisa, Türkiye (email: saira.rafique@vestel.com.tr).}
\thanks{Ahmed Naeem is with the School of Electrical Engineering and Computer Science National University of Sciences and Technology, Islamabad, Pakistan (email: ahmed.naeem@seecs.edu.pk).}
\thanks{Chong Han is with the Terahertz Wireless Communications Interdisciplinary Research Center (TWC IRC) and also the Cooperative Medianet Innovation Center (CMIC), School of Information Science and Electronic Engineering, Shanghai Jiao Tong University, Shanghai 200240,
    China (e-mail: chong.han@sjtu.edu.cn).}
\thanks{Hüseyin Arslan is with the Department of Electrical and Electronics Engineering, Istanbul Medipol University, Istanbul, 34810, Türkiye (email: huseyinarslan@medipol.edu.tr).}}
\markboth{}%
{Shell \MakeLowercase{\textit{et al.}}: A Sample Article Using IEEEtran.cls for IEEE Journals}
\maketitle

\begin{abstract}
High pulse repetition frequency (PRF) operation improves Doppler unambiguous sensing but reduces the maximum unambiguous range of a terahertz (THz) pulsed radar. Echoes from targets beyond this range can be incorrectly associated with subsequent transmitted pulses, causing the targets to appear at shorter ranges. This paper proposes a receiver-side framework that exploits the THz-specific temporal broadening effect as a distance-dependent signature to resolve range ambiguity. For each detected echo, the receiver applies broadening-matched filtering with references corresponding to candidate target ranges. The filter outputs are combined across pulses with Doppler compensation, and the largest normalised combined response identifies the ambiguity order needed to recover the true target range. The selected reference is then reused to reduce mismatch loss in subsequent range and velocity estimation. Simulations under known atmospheric conditions, with targets occupying distinct apparent-delay bins, demonstrate reliable range recovery beyond the conventional unambiguous interval. Range estimation errors conditioned on correct ambiguity-order identification, together with the reported velocity estimation errors, approach their corresponding derived Cramér--Rao bounds. The method preserves a fixed transmitted pulse shape, PRF, and sampling rate, at the expense of a higher signal-to-clutter-plus-noise ratio requirement for ambiguity identification relative to the evaluated diversity-based techniques.
\end{abstract}

\section{Introduction}
\IEEEPARstart{T}{erahertz} (THz) band (0.1–10 THz), with its wide bandwidths and millimeter-to-micrometer wavelengths, offers considerable potential for high-resolution sensing in future wireless systems. These properties support localization, imaging, object tracking, and environmental mapping for applications such as intelligent transportation, industrial automation, biomedical inspection, and security screening  \cite{11075517, AKYILDIZ201416}. THz sensing also offers opportunities for environmental perception in embodied AI, motivating its investigation for robotic platforms, including humanoid robots. However, harnessing these sensing capabilities requires accounting for unique THz channel characteristics, including molecular absorption (MoA), temporal broadening, and surface-dependent reflection and scattering \cite{11075517}. Atmospheric molecular resonances introduce frequency-selective attenuation and dispersion. Attenuation changes the relative amplitudes of the signal’s frequency components, whereas dispersion introduces differences in their group delays, which can spread a pulse in time even along a single line-of-sight path \cite{10707074}. This increase in the pulse’s temporal duration is referred to as the temporal broadening effect (TBE). The resulting pulse shape depends on the propagation distance, atmospheric conditions, and the transmitted pulse’s spectral amplitude and phase. Although broadening can impair reception, its distance dependence provides a potential signature for distinguishing echoes beyond their arrival delays.

\par Existing studies demonstrate how THz propagation dependencies can be incorporated into signal processing. Distance-aware multicarrier modulation adapts transmission to distance-dependent
spectral windows \cite{6884190}, while distance-adaptive absorption-peak modulation exploits molecular attenuation for covert communication \cite{9271892}. Temporal broadening has also been exploited in artificial-noise-based secure communication and pulse-width adaptation for inter-symbol interference mitigation and energy efficiency \cite{9497766,11195826}. Extending this principle to joint sensing and communication, TBE-aware multiplexing embeds low-power communication pulses for a nearby user within sensing guard intervals allocated to accommodate broadening at a farther sensing receiver \cite{11692668}. These studies demonstrate the utility of distance-dependent propagation effects and motivate investigating their use for range-ambiguity identification in THz sensing.

\par Pulsed systems have been investigated for THz sensing and communication. Customized on--off keying supports simultaneous THz imaging, information transfer, and power transfer \cite{10136710}. Experimental pulsed operation is also demonstrated by the THz time-domain spectroscopy (THz-TDS)-based radar in \cite{11320100}, where a high source repetition rate enables compact optical delay hardware but limits the native unambiguous ranging interval. For coherent pulsed radar, increasing the pulse repetition frequency (PRF) extends the Doppler interval that can be measured without aliasing \cite{11357567}. Shorter wavelengths produce larger Doppler shifts for a given radial velocity; therefore, maintaining the same unambiguous velocity interval requires a higher PRF as the wavelength decreases. Increasing the PRF also permits more pulses within a fixed coherent processing interval (CPI). At fixed per-pulse energy, coherent integration can consequently accumulate more signal energy, with a corresponding increase in average transmit power. These considerations motivate the high-PRF THz pulsed radar considered in this work. A shorter pulse repetition interval (PRI), however, reduces the maximum unambiguous range: the distance corresponding to a round-trip propagation delay of one PRI. Echoes from targets beyond this range return after subsequent pulses have been transmitted. When an echo is associated with a later transmission instead of its originating pulse, the receiver assigns it a shorter apparent range. Different physical ranges, separated by integer multiples of the maximum unambiguous range, can therefore share the same apparent delay. The corresponding integer is the ambiguity order. This ambiguity results from periodic transmission and exists independently of temporal broadening or overlap between received echoes. Resolving it requires additional information that distinguishes the candidate physical ranges.

\subsection{Related Works}
Conventional radar methods address range ambiguity through multiple or staggered pulse repetition frequencies (PRFs), waveform diversity, and receiver-side sampling diversity. Multiple or staggered PRFs change the range-folding relationship across observations, while OFDM waveform diversity and pulse-diverse processing introduce transmission-dependent signatures for ambiguity mitigation \cite{6203367,5960555}. Receiver-side sampling diversity instead varies the effective sampling rate across PRIs and uses matched-filter gain and sidelobe characteristics to identify ambiguous hyperbolic frequency modulated (HFM) echoes \cite{11357567}. These approaches were developed outside the THz setting and provide relevant comparison methods.
Several approaches have been proposed to extend unambiguous ranging in THz sensing systems. Customized on--off keying uses data-dependent pulse timing to identify ranges consistent across transmissions \cite{10136710}. Stepped-frequency optical intensity modulation extends the native unambiguous ranging interval of THz-TDS-based radar \cite{11320100}, while photonic THz chaotic signals enable unambiguous ranging through a different signal-generation approach \cite{deng2025photonic}. Propagation-aware processing offers a complementary direction. For instance, frequency-selective THz channel modeling improves range estimation compared with a flat-channel assumption \cite{9827896}. Building on this observation, the present work identifies the PRI ambiguity order through normalised correlations with propagation-matched references and coherent Doppler-compensated integration. The receiver retains a fixed transmitted pulse shape, PRF, and sampling rate; its identification performance depends on reference separation, atmospheric-model accuracy, and signal-to-clutter-plus-noise ratio (SCNR).

\subsection{Contributions and Organization}
This paper develops a receiver-side framework to identify and resolve range ambiguity in monostatic THz pulsed radar using distance-dependent temporal broadening. The framework assumes known atmospheric conditions and targets occupying distinct apparent-delay bins. Its main contributions are summarized as follows:
\begin{itemize}
\item A receiver-side framework is proposed to resolve range ambiguity using distance-dependent THz-specific TBE. Normalised, Doppler-compensated coherent correlations with candidate propagation-matched references identify the ambiguity order. The selected reference is reused for delay and Doppler estimation and range reconstruction, while the transmitted pulse shape, PRF, and sampling rate remain fixed.

\item An analytical framework distinguishes detection, ambiguity identification, and estimation accuracy. It quantifies transmit-reference mismatch loss, derives local range and velocity Cramér--Rao bounds with known ambiguity order, and characterizes the role of candidate-reference similarity in ambiguity discrimination. An optional scalar classifier is formulated for intervals with a strictly increasing, sufficiently separated calibrated broadening response.
\item Simulation results demonstrate range recovery beyond the conventional unambiguous interval and improved correlation-peak quality through propagation-matched processing. Reliable ambiguity-order identification requires higher SCNR than echo detection, representing the cost of resolving range ambiguity while retaining a fixed transmitted pulse shape, PRF, and sampling rate.
\end{itemize}

The remainder of this paper is organized as follows. Section~II presents the THz pulsed radar signal model, the MoA channel, and the resulting broadening signature. Section~III formulates the received signal and the folding mechanism. Section~IV introduces the proposed identification and elimination algorithm. Section~V analyses detection, peak quality, estimation bounds, and identification reliability. Section~VI presents the simulation setup and numerical results. Finally, Section~VII concludes the paper.

\section{System Model}\label{sec:system_model}
\begin{figure}[h!] 
\centering
\resizebox{1\columnwidth}{!}{
\includegraphics{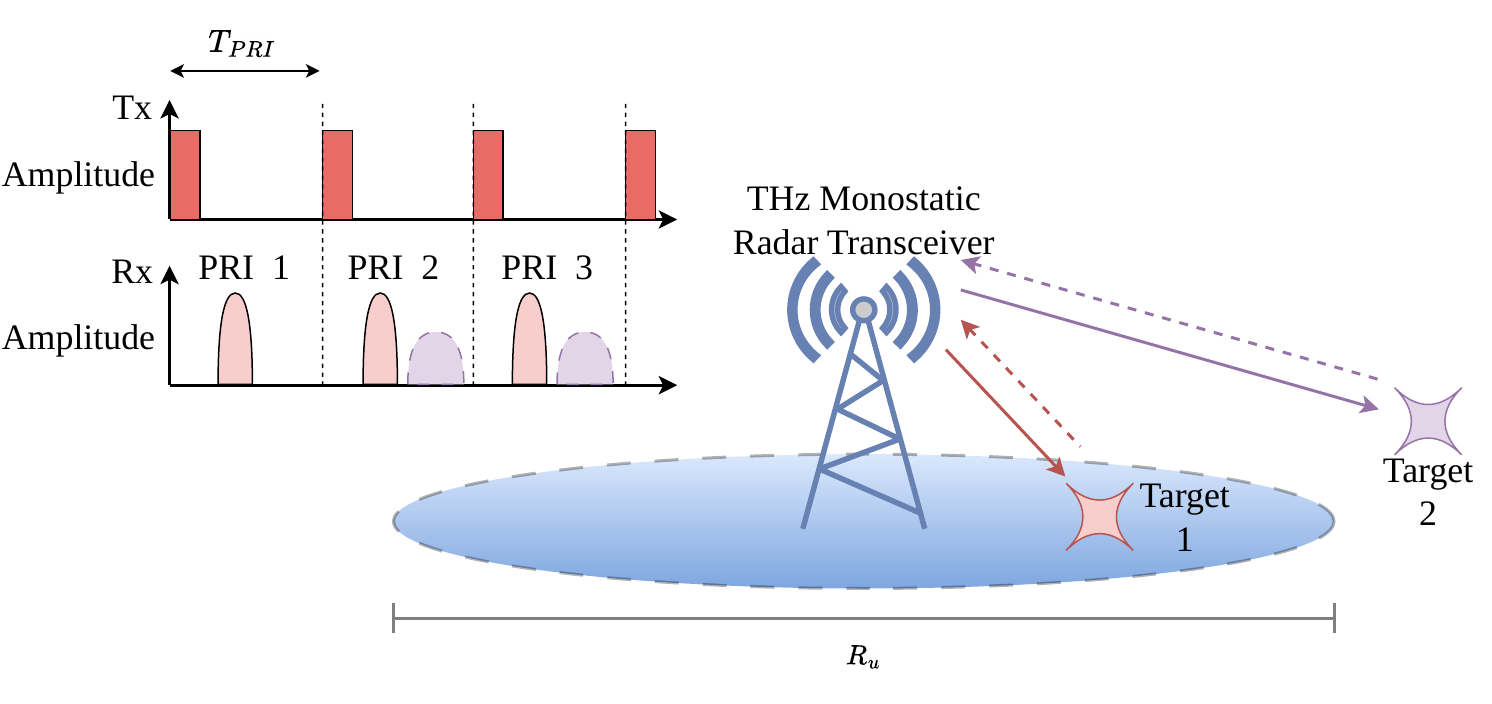}}
\caption{System model of the proposed THz monostatic pulsed radar with TBE-aware range ambiguity identification.}
\label{sys}
\end{figure}

Consider a monostatic THz pulsed radar transceiver as shown in Fig.~\ref{sys}, observed over a CPI containing $P$ receive PRIs. Periodic transmission is established before the CPI. The corresponding $P$-pulse transmitted signal segment is given by
\begin{equation}\label{eq:tx_pulse_train}
    s(t) \;=\; \sum_{p=0}^{P-1} g\!\big(t - p\, T_{\mathrm{PRI}}\big),
\end{equation}
where $T_{\mathrm{PRI}}$ is the pulse repetition interval (PRI) with the corresponding pulse repetition frequency 
$f_{\mathrm{PRF}} = 1/T_{\mathrm{PRI}}$, and $g(t)$ is the baseband transmit pulse. For analytical tractability and 
hardware simplicity, a rectangular pulse of duration $T_p$ is adopted, expressed as
\begin{equation}\label{eq:rect_pulse}
g_0(t)=\mathrm{rect}\!\left(\frac{t}{T_p}\right).
\end{equation}
The ideal pulse spectrum and the filtered transmit spectrum are
\begin{equation}\label{eq:rect_fourier}
\begin{aligned}
G_0(\nu)&=T_p\,\mathrm{sinc}(\nu T_p),\\
G(\nu)&=G_0(\nu)A_{\mathrm{tx}}(\nu),
\end{aligned}
\end{equation}
where $\mathrm{sinc}(x)=\sin(\pi x)/(\pi x)$, $\nu$ is the baseband frequency offset, and $A_{\mathrm{tx}}(\nu)$ is the transmit-filter response. The transmitted pulse is $g(t)=\mathcal F^{-1}_{\nu\rightarrow t}\{G(\nu)\}$.
The filter confines the spectrum to $|\nu|\leq B/2$. Here, $T_p$ is the nominal rectangular-pulse duration,
and $B\approx1/T_p$ is the adopted bandwidth choice; the filtered pulse is not strictly time-limited.

The radar scene comprises $L$ point reflectors representing humans or other static or slow-moving obstacles within the line-of-sight (LoS) sensing region. The $\ell$th target ($\ell \in \{1, 2, \ldots, L\}$) is at range $R_\ell$, with corresponding round-trip propagation distance $d_\ell = 2R_\ell$ and two-way delay $\tau_\ell = 2R_\ell/c$, where $c = 3 \times 10^8$~m/s. The maximum unambiguous range of the radar is $R_u \;=\; \frac{c\, T_{\mathrm{PRI}}}{2}$. Targets at ranges $R_\ell \geq R_u$ produce echoes that arrive after one or more subsequent pulses have been 
transmitted, and are observed at the apparent folded range $R_{\mathrm{app},\ell} = R_\ell - q_\ell R_u$, where 
$q_\ell = \lfloor R_\ell / R_u \rfloor \in \{0, 1, 2, \ldots\}$ is the ambiguity order.
THz propagation is predominantly LoS due to the strong reflection, scattering, and diffraction losses experienced by THz waves, which render the energy carried by non-line-of-sight (NLoS) multipath components negligible relative to the LoS component \cite{9271892,9497766}. Highly directional THz transceivers are adopted to overcome the elevated THz path loss \cite{6884190}, and the dominant LoS path is therefore used to model the round-trip THz channel between radar and each target.

\subsection{THz Channel Model}\label{subsec:thz_channel}
A homogeneous atmosphere with known temperature, pressure, and gas composition is assumed. For target $\ell$, the round-trip propagation distance is $d_\ell=2R_\ell$, and the delay-free channel response is
\begin{equation}\label{eq:delay_free_channel}
\tilde H_\ell(f) = \tilde H(f;d_\ell)= H_{\mathrm{abs}}(f;d_\ell)e^{-j\psi(f;d_\ell)},
\end{equation}
where $f$ denotes absolute frequency. The baseband offset is $\nu=f-f_c$, and the transmitted spectrum is
$G(f-f_c)$. The MoA and dispersive phase are included in $\tilde H_\ell(f)$, while spreading loss, target reflectivity, antenna gains, and constant phase are absorbed into $\alpha_\ell$ and approximated as constant over the occupied band. The field attenuation follows the Beer--Lambert law
\cite{5995306,jornet2010channel} and is expressed as
\begin{equation}\label{eq:expoterm}
H_{\mathrm{abs}}(f;d_\ell)
= \exp\!\left[-\frac{k(f)d_\ell}{2}\right],
\end{equation}
where $k(f)$ is the power absorption coefficient in $\mathrm{m^{-1}}$. The factor $1/2$ converts power
attenuation to field attenuation. For a causal propagation model whose refractive index
approaches unity at infinite frequency, absorption and dispersion are related by the Kramers--Kronig relation \cite{10707074} as
\begin{equation}\label{eq:phase}
n_{\mathrm r}(f)
= 1+\frac{c}{2\pi^2}\operatorname{PV}
\int_0^\infty
\frac{k(f')}{f'^2-f^2}\,\mathrm df',
\end{equation}
where $\operatorname{PV}$ denotes the Cauchy principal value. This relation involves the broadband absorption spectrum, including frequencies outside the radar band. For numerical evaluation, the channel uses the oxygen and water-vapour line parameters and dry-air continuum model of ITU-R P.676-12 \cite{series2019attenuation}. The imaginary and real parts of the implemented complex refractivity
determine $k(f)$ and $n_{\mathrm r}(f)-1$, respectively. The excess propagation phase relative to vacuum is
\begin{equation}\label{eq:psi}
\psi(f;d_\ell)
= \frac{2\pi f[n_{\mathrm r}(f)-1]d_\ell}{c}.
\end{equation}
The vacuum delay $\tau_\ell=d_\ell/c$ is represented explicitly in the received-signal model and excluded
from $\tilde H_\ell(f)$; its constant carrier phase is absorbed into $\alpha_\ell$. The excess group delay is
\begin{equation}\label{eq:excess_group_delay}
\tau_{\mathrm{ex}}(f;d_\ell)
= \frac{1}{2\pi}
\frac{\partial\psi(f;d_\ell)}{\partial f},
\end{equation}
and the local phase curvature is
\begin{equation}\label{eq:phase_curvature}
\Psi_2(d_\ell)
= \left.
\frac{\partial^2\psi}{\partial\omega^2}
\right|_{\omega=2\pi f_c},
\end{equation}
where $\omega=2\pi f$. Under fixed atmospheric conditions, these quantities scale linearly with $d_\ell$.
Frequency-dependent attenuation and group delay reshape the received pulse. The delay-free baseband impulse response is
\begin{equation}\label{eq:baseband_channel}
\tilde h_\ell(t)
= \mathcal F^{-1}_{\nu\rightarrow t}
\left\{\tilde H_\ell(f_c+\nu)\right\}.
\end{equation}

\subsection{Temporal Broadening Effect}\label{subsec:tbe}
Let $\tilde h_\ell(t)$ denote the delay-free baseband impulse response associated with $\tilde H_\ell(f_c+\nu)$, where $\nu$ is the baseband frequency offset. The noise-free single-pulse echo from target $\ell$ is
\begin{equation}\label{eq:tbe_conv}
s_{\mathrm r,\ell}(t) = \alpha_\ell (g*\tilde h_\ell)(t-\tau_\ell),
\end{equation}
where $\alpha_\ell$ accounts for target reflectivity, antenna gains, spreading loss, and constant phase. Frequency-dependent attenuation and group delay reshape the echo. Under fixed atmospheric conditions, this distortion depends on the round-trip distance $d_\ell$. Temporal broadening is characterized using a central interval containing a fraction $\varrho\in(0,1)$ of the signal energy, with equal energy in the excluded tails. For a nonzero finite-energy signal $x(t)$, the normalised cumulative energy is
\begin{equation}\label{eq:cum_energy}
F_x(t)
=
\frac{\displaystyle\int_{-\infty}^{t}|x(u)|^2\,\mathrm du}
     {\displaystyle\int_{-\infty}^{\infty}|x(u)|^2\,\mathrm du}.
\end{equation}
The corresponding duration is
\begin{equation}\label{eq:pulse_duration}
D(x)
=
F_x^{-1}\!\left(\frac{1+\varrho}{2}\right)
-
F_x^{-1}\!\left(\frac{1-\varrho}{2}\right),
\end{equation}
where $F_x^{-1}(a)=\inf\{t:F_x(t)\geq a\}$. Broadening is evaluated at the output of the filter matched to the transmitted pulse. With $g^-(t)=g^*(-t)$, the reference and propagated responses are
\begin{equation}\label{eq:tbe_filter_responses}
z_g=g*g^-,
\qquad
z_{d_\ell}=(g*\tilde h_\ell)*g^-.
\end{equation}
The broadening factor is
\begin{equation}\label{eq:beta}
\beta(d_\ell)
=
\frac{D(z_{d_\ell})}{D(z_g)},
\end{equation}
with reference duration $D_0=D(z_g)$. Neither monotonicity nor $\beta(d)\geq1$ is assumed. For the optional scalar classifier, $\beta_{\mathrm{cal}}(d)$ is the sampled propagated-response duration normalised by $D_{0,\mathrm s}>0$. Calibration uses the receiver's sampling rate, matched filter, energy fraction, finite-window convention, and sample-index quantiles, accounting for truncation and sampling alignment. Scalar classification requires a strictly increasing calibrated map with separation exceeding measurement and calibration errors; the main correlation-based receiver requires no monotonicity.

\section{Received Signal Model}\label{sec:received_signal}
The received signal comprises target echoes, clutter, and additive noise. Periodic transmission is established
before the observed CPI, which contains $P$ receive PRIs. For target $\ell$, the echo generated by the $p$th
transmitted pulse in \eqref{eq:tx_pulse_train} is given by
\begin{equation}\label{eq:per_pulse_echo}
s_{\mathrm r,\ell,p}(t) = \alpha_\ell e^{-j2\pi f_{D,\ell}pT_{\mathrm{PRI}}}
(g*\tilde h_\ell) \big(t-pT_{\mathrm{PRI}}-\tau_\ell\big),
\end{equation}
where $p=0,\ldots,P-1$ and $f_{D,\ell}=2v_\ell f_c/c$. The exponential specifies the slow-time Doppler phase
convention. Doppler phase variation within each received echo is neglected, requiring
$|f_{D,\ell}|T_{\mathrm{echo},\ell}\ll1$, where $T_{\mathrm{echo},\ell}$ is its effective duration.
Target amplitudes, delays, Doppler frequencies, and propagation responses are approximately constant
over the CPI. Over the observed interval $0\leq t<PT_{\mathrm{PRI}}$, the composite baseband signal is given by
\begin{equation}\label{eq:received_signal_full}
y(t) = \sum_{p=0}^{P-1}\sum_{\ell=1}^{L} s_{\mathrm r,\ell,p}(t)
+y_{\mathrm{pre}}(t)+c(t)+w(t),
\end{equation}
where $y_{\mathrm{pre}}(t)$ comprises target echoes generated by transmissions preceding the
$P$-pulse segment, $c(t)$ is the composite clutter return, and $w(t)$ is additive receiver noise.
Including $y_{\mathrm{pre}}(t)$ accounts for earlier transmissions contributing to the initial receive PRIs. The sampled CPI noise vector satisfies $\mathbf w\sim\mathcal{CN}(\mathbf0,\mathbf C_w)$, where $\mathbf C_w$ accounts for receive filtering and sampling. Under the white-sample approximation,
$\mathbf C_w=\sigma_w^2\mathbf I$. Clutter is modeled as a superposition of returns from Poisson-distributed scatterers within the modeled scene. Scatterer locations and Swerling-I reflectivity realizations remain fixed within each CPI, while their phases evolve according to their Doppler shifts.
For target $\ell$, the ambiguity order $q_\ell=\lfloor R_\ell/R_u\rfloor$ counts the complete PRIs in its geometric round-trip delay. The nominal arrival time of transmitted pulse $p$ is $pT_{\mathrm{PRI}}+\tau_\ell$, corresponding to receive PRI $p+q_\ell$. The geometric apparent delay is
\begin{equation}\label{eq:tau_app}
\tau_{\mathrm{app},\ell}= \tau_\ell-q_\ell T_{\mathrm{PRI}},
\qquad 0\leq\tau_{\mathrm{app},\ell}<T_{\mathrm{PRI}},
\end{equation}
and the corresponding apparent range is
\begin{equation}\label{eq:R_app}
R_{\mathrm{app},\ell} =
\frac{c\tau_{\mathrm{app},\ell}}{2} = R_\ell-q_\ell R_u,
\qquad 0\leq R_{\mathrm{app},\ell}<R_u.
\end{equation}

An echo received in PRI $p$ originates from pulse $p-q_\ell$. The constant Doppler factor $e^{j2\pi f_{D,\ell}q_\ell T_{\mathrm{PRI}}}$ is absorbed
into the unknown complex amplitude, preserving the slow-time phase progression $e^{-j2\pi f_{D,\ell}pT_{\mathrm{PRI}}}$. Excess channel group delay can shift the transmit-reference correlation peak from the geometric apparent delay, so this peak supplies an initial estimate for propagation-matched
refinement. Targets are assumed resolved in apparent delay, with negligible neighboring-pulse tails and PRI-edge truncation. Delay-only processing cannot distinguish physical ranges separated by integer multiples of $R_u$.

\section{Proposed Approach}\label{sec:proposed}
The proposed framework identifies the ambiguity order by comparing each detected echo with propagation-matched references for candidate physical ranges. The complex correlation outputs are combined coherently across the CPI with Doppler compensation, and the largest normalised coherent correlation statistic determines the selected order. The selected reference is
reused for delay and Doppler estimation to reconstruct
physical range and radial velocity.
\subsection{Initial Detection and Broadening Estimation}\label{subsec:tbe_evaluation}
The receive CPI contains $P$ PRIs, indexed by $p=0,\ldots,P-1$. The Doppler search uses a specified grid $\mathcal F_D\subset [-f_{\mathrm{PRF}}/2,f_{\mathrm{PRF}}/2)$. Sampling at $f_s=2B$ gives
\begin{equation}\label{eq:per_pri_window}
y_p[n] = y\!\left(pT_{\mathrm{PRI}}+\frac{n}{f_s}\right), \quad n=0,\ldots,N_{\mathrm{ft}}-1,
\end{equation}
where $N_{\mathrm{ft}}=T_{\mathrm{PRI}}f_s\in\mathbb N$. With $g[n]=g(n/f_s)$, the transmit-reference correlation is
\begin{equation}\label{eq:mf_output}
z_p[k]=\sum_{n=0}^{N_{\mathrm{ft}}-1}y_p[n]g^*[n-k], \quad k=0,\ldots,N_{\mathrm{ft}}-1.
\end{equation}
Both positive and negative reference indices are evaluated from the centered pulse $g(t)$. The full-CPI detection statistic is expressed as
\begin{equation}\label{eq:detection_statistic}
\mathcal T(k,f_D)=
\left|\sum_{p=0}^{P-1}
z_p[k]e^{j2\pi f_DpT_{\mathrm{PRI}}}\right|^2,
\qquad f_D\in\mathcal F_D.
\end{equation}
Resolved local maxima of
$\max_{f_D\in\mathcal F_D}\mathcal T(k,f_D)$
exceeding $\eta_{\mathrm{det}}$ are retained.
The threshold is set under the adopted noise and clutter model for the prescribed per-cell false-alarm
probability. For detection $\ell$ at CPI peak index $k_\ell^{\mathrm{init}}$, the initial apparent-delay estimate is written as
\begin{equation}\label{eq:initial_delay}
\hat\tau_{\mathrm{app},\ell}^{\mathrm{init}} =
\frac{k_\ell^{\mathrm{init}}}{f_s}.
\end{equation}
Channel-induced correlation-peak displacement can bias this estimate relative to the geometric apparent delay. For per-PRI broadening measurements, $k_\ell^{(p)}$ is the associated local correlation peak in the detected echo's delay neighborhood. The corresponding apparent-delay estimate is
\begin{equation}\label{eq:apparent_estimates}
\hat\tau_{\mathrm{app},\ell}^{(p)} = \frac{k_\ell^{(p)}}{f_s}.
\end{equation}
A window $\mathcal W_{\ell,p}$ contains the selected response and excludes neighboring resolved echoes.
For duration estimation, the squared magnitude is adjusted using the estimated clutter and noise power as
\begin{equation}\label{eq:noise_sub_energy}
\tilde E_{\ell,p}[k] = \max\{|z_p[k]|^2-\hat\sigma_u^2,0\}, \qquad k\in\mathcal W_{\ell,p}.
\end{equation}
Here, $\hat\sigma_u^2$ estimates the mean clutter and noise power at the matched-filter output
from target-free reference cells. For positive total residual energy,
\begin{equation}\label{eq:per_pri_cum}
\hat F_{\ell,p}[m] = \frac{ \sum_{\substack{k\in\mathcal W_{\ell,p}\\k\leq m}} \tilde E_{\ell,p}[k]}
{\sum_{k\in\mathcal W_{\ell,p}} \tilde E_{\ell,p}[k]}.
\end{equation}
The estimated duration is
\begin{equation}\label{eq:per_pri_duration}
\hat D_\ell^{(p)} = \frac{ \hat F_{\ell,p}^{-1}((1+\varrho)/2)
- \hat F_{\ell,p}^{-1}((1-\varrho)/2)}{f_s},
\end{equation}
where the inverse returns the first sample index reaching the requested cumulative-energy fraction. The per-PRI broadening estimate is
\begin{equation}\label{eq:beta_hat}
\hat\beta_\ell^{(p)}= \frac{\hat D_\ell^{(p)}}{D_{0,\mathrm s}}.
\end{equation}
Measurements with zero total residual energy are undecided. Subtracting the estimated clutter and noise power does not guarantee an unbiased duration estimate. These broadening measurements are used only by the interval classifier in Section~\ref{subsubsec:interval}.

\subsection{Range Disambiguation} \label{subsec:range_disambiguation}
The receiver tests $q=0,\ldots,q_{\max}$, with $q_\ell\leq q_{\max}$ assumed for all modeled targets.
\subsubsection{Propagation-Matched References}\label{subsubsec:matched}
For candidate order $q$ and apparent delay $\tau$, the round-trip distance is
\begin{equation}\label{eq:candidate_distance}
d_q(\tau)=c\tau+2qR_u.
\end{equation}
The corresponding delay-free reference spectrum is
\begin{equation}\label{eq:template_q}
T_q(\nu;\tau)= G(\nu)\tilde H(f_c+\nu;d_q(\tau)).
\end{equation}
Here, $\tilde H(f;d)$ denotes the delay-free propagation response at round-trip distance $d$, with
$\tilde H_\ell(f)=\tilde H(f;d_\ell)$.
The aligned sampled reference is
\begin{equation}\label{eq:template_samples}
\mathbf t_q(\tau)=\left[\left(\mathcal F^{-1}_{\nu\rightarrow t}\{T_q(\nu;\tau)\}\right)
\left(\frac{n}{f_s}-\tau\right)\right]_{n=0}^{N_{\mathrm{ft}}-1}.
\end{equation}

The received PRI vector is $\mathbf y_p=[y_p[n]]_{n=0}^{N_{\mathrm{ft}}-1}$.
Both $\mathbf t_q(\tau)$ and $\mathbf y_p$ are column vectors in $\mathbb C^{N_{\mathrm{ft}}}$, using the
same time origin and observation interval. Their complex correlation is
\begin{equation}\label{eq:lambda_q}
x_\ell^{(p)}(q;\tau)=\mathbf t_q^H(\tau)\mathbf y_p.
\end{equation}
For order identification, every candidate reference is constructed at $\tau=\hat\tau_{\mathrm{app},\ell}^{\mathrm{init}}$ and remains fixed across the CPI.

\subsubsection{Coherent Combining}
\label{subsubsec:combining}
All $P$ received PRIs are combined coherently. The target amplitude, apparent delay, and propagation
response are assumed approximately constant over the CPI, while the modeled slow-time Doppler phase is retained. The nonzero received energy is given by
\begin{equation}\label{eq:block_energy}
E=\sum_{p=0}^{P-1}\|\mathbf y_p\|^2,
\end{equation}
and the normalised coherent correlation statistic is
\begin{equation}\label{eq:block_stat}
S_\ell(q;\tau,f_D) = \frac{\left|\sum_{p=0}^{P-1}x_\ell^{(p)}(q;\tau)e^{j2\pi f_DpT_{\mathrm{PRI}}}\right|^2}
{E\|\mathbf t_q(\tau)\|^2}.
\end{equation}
Only references satisfying$\|\mathbf t_q(\tau)\|^2>0$ are considered. The positive exponential compensates the modeled negative slow-time Doppler phase at the matching frequency. Maximization over $\mathcal F_D$ requires
no prior Doppler estimate. The selected ambiguity order is
\begin{equation}\label{eq:q_hat_accum}
\hat q_\ell=\arg\max_{q=0,\ldots,q_{\max}}\max_{f_D\in\mathcal F_D}S_\ell\!\left( q;\hat\tau_{\mathrm{app},\ell}^{\mathrm{init}},f_D
\right).
\end{equation}
The initial apparent-delay estimate is held fixed during order selection. Its estimation error,
including channel-induced correlation-peak displacement, can affect the selected order.
Subsequent estimation retains $\hat q_\ell$. The statistic is invariant to nonzero scalar scaling
of the received record and each reference, with $0\leq S_\ell\leq P$.
Identification remains sensitive to SCNR, reference similarity, alignment errors, and channel-model errors. Clutter returns can alter the candidate correlation statistics and lead to incorrect ambiguity-order decisions.

\subsubsection{Broadening-Based Interval Classification}
\label{subsubsec:interval}
As an alternative to propagation-matched order selection, the per-PRI broadening estimate can be
classified using a strictly increasing sampled calibration over $[0,2(q_{\max}+1)R_u]$. The thresholds are defined as
\begin{equation}\label{eq:beta_thresholds}
b_q=\beta_{\mathrm{cal}}(2qR_u),
\qquad q=0,\ldots,q_{\max}+1.
\end{equation}
The per-PRI decision is
\begin{equation}\label{eq:q_hat_pri}
\hat q_{\ell,\mathrm{int}}^{(p)}=
\begin{cases}q,&b_q\leq\hat\beta_\ell^{(p)}<b_{q+1},
\\
\text{undecided},&\text{otherwise}.
\end{cases}
\end{equation}
Here, $q\in\{0,\ldots,q_{\max}\}$. This scalar classifier is not used for the reported
numerical results, which use \eqref{eq:q_hat_accum}.

\subsection{Matched-Reference Delay and Doppler Estimation}
\label{subsec:doppler}
After selecting $\hat q_\ell$, the receiver reuses the corresponding propagation-matched reference.
Its delay-free spectrum is
\begin{equation}\label{eq:selected_reference_spectrum}
\widehat T_\ell(\nu)
=
T_{\hat q_\ell}\!
\left(\nu;\hat\tau_{\mathrm{app},\ell}^{\mathrm{init}}\right).
\end{equation}
The propagation response in $\widehat T_\ell(\nu)$ remains fixed during subsequent delay and Doppler estimation. For trial apparent delay $\tau$, the shifted sampled reference is
\begin{equation}\label{eq:selected_reference_samples}
\widehat{\mathbf t}_\ell(\tau) = \left[\left(\mathcal F^{-1}_{\nu\rightarrow t}\{\widehat T_\ell(\nu)\}\right)\left(\frac{n}{f_s}-\tau\right)\right]_{n=0}^{N_{\mathrm{ft}}-1}.
\end{equation}
The corresponding correlation output is
\begin{equation}\label{eq:selected_reference_correlation}
\widehat z_{\ell,p}(\tau)=\widehat{\mathbf t}_\ell^H(\tau)\mathbf y_p.
\end{equation}
The local delay search interval$\mathcal D_\ell\subset[0,T_{\mathrm{PRI}})$ accounts for initial-delay uncertainty and channel-induced correlation-peak displacement, while excluding neighboring resolved responses.
The per-PRI refined apparent-delay estimate is
\begin{equation}\label{eq:matched_delay_estimate}
\widehat\tau_{\ell,\mathrm m}^{(p)}=\arg\max_{\tau\in\mathcal D_\ell} \left|\widehat z_{\ell,p}(\tau)\right|^2.
\end{equation}
The CPI apparent-delay estimate is
\begin{equation}\label{eq:refined_apparent_delay}
\hat\tau_{\mathrm{app},\ell}=\frac{1}{P}\sum_{p=0}^{P-1}\widehat\tau_{\ell,\mathrm m}^{(p)}.
\end{equation}

The matched-filter outputs at the associated per-PRI peaks form the slow-time sequence
\begin{equation}\label{eq:matched_slow_time}
u_\ell[p]=\widehat z_{\ell,p}\!\left(\widehat\tau_{\ell,\mathrm m}^{(p)}\right),\qquad p=0,\ldots,P-1.
\end{equation}
The Doppler estimate is
\begin{equation}\label{eq:doppler_estimate}
\hat f_{D,\ell}=\arg\max_{f_D\in\mathcal F_D}\left|\sum_{p=0}^{P-1}u_\ell[p]e^{j2\pi f_DpT_{\mathrm{PRI}}}\right|^2.
\end{equation}
The physical range and radial velocity are
\begin{equation}\label{eq:true_range_reconstruction}
\hat R_\ell=\frac{c\hat\tau_{\mathrm{app},\ell}}{2}+\hat q_\ell R_u,
\end{equation}
\begin{equation}\label{eq:velocity_estimate}
\hat v_\ell=\frac{c\hat f_{D,\ell}}{2f_c}.
\end{equation}
Propagation-matched filtering reduces transmit-reference mismatch. Estimation accuracy remains affected by noise, clutter, peak-selection errors, channel-model errors, and incorrect ambiguity-order decisions. The selected propagation distance is not updated during refinement, so an error in its initial value can leave residual reference mismatch. The proposed technique is summarized in Algorithm \ref{alg:tbe_disambig}.
\begin{algorithm}[t!]
\caption{TBE-Aware Range Ambiguity Identification and Elimination}
\label{alg:tbe_disambig}
\begin{algorithmic}[1]

\STATE \textbf{Initialization:}
Specify $q_{\max}$, $\mathcal F_D$, and the
propagation model for the known atmospheric conditions.
\STATE Sample $y(t)$ at $f_s=2B$ and form
$\{\mathbf y_p\}_{p=0}^{P-1}$
using \eqref{eq:per_pri_window}.
\STATE Compute $E$ using \eqref{eq:block_energy}.
\IF{$E=0$}
    \STATE \textbf{return} $\varnothing$.
\ENDIF

\FOR{$p=0,\ldots,P-1$}
    \STATE Compute $z_p[k]$ using \eqref{eq:mf_output}.
\ENDFOR
\STATE Form $\mathcal T(k,f_D)$ using
\eqref{eq:detection_statistic}.
\STATE Set $\eta_{\mathrm{det}}$ under the adopted
noise and clutter model for the prescribed
per-cell false-alarm probability.
\STATE Retain resolved local maxima of
$\max_{f_D\in\mathcal F_D}\mathcal T(k,f_D)$
exceeding $\eta_{\mathrm{det}}$.
\STATE Obtain $\hat L$ detections and their
initial apparent-delay estimates using
\eqref{eq:initial_delay}.

\FOR{$\ell=1,\ldots,\hat L$}
    \FOR{$q=0,\ldots,q_{\max}$}
        \STATE Construct
        $\mathbf t_q(
        \hat\tau_{\mathrm{app},\ell}^{\mathrm{init}})$
        using \eqref{eq:candidate_distance}--%
        \eqref{eq:template_samples}.
        \STATE For a reference with positive squared
        norm, compute the per-PRI correlations using
        \eqref{eq:lambda_q} and evaluate
        \eqref{eq:block_stat} over $\mathcal F_D$.
    \ENDFOR
    \STATE Select $\hat q_\ell$ using
    \eqref{eq:q_hat_accum}, holding the initial
    apparent-delay estimate fixed.

    \STATE Retain $\widehat T_\ell(\nu)$ using
    \eqref{eq:selected_reference_spectrum}.
    \STATE Select the local delay interval
    $\mathcal D_\ell$.
    \FOR{$p=0,\ldots,P-1$}
        \STATE Shift the retained reference and
        compute $\widehat z_{\ell,p}(\tau)$ using
        \eqref{eq:selected_reference_samples} and
        \eqref{eq:selected_reference_correlation}.
        \STATE Estimate
        $\widehat\tau_{\ell,\mathrm m}^{(p)}$
        using \eqref{eq:matched_delay_estimate}.
    \ENDFOR
    \STATE Compute $\hat\tau_{\mathrm{app},\ell}$
    using \eqref{eq:refined_apparent_delay}.

    \STATE Form $u_\ell[p]$ using
    \eqref{eq:matched_slow_time}.
    \STATE Estimate $\hat f_{D,\ell}$ using
    \eqref{eq:doppler_estimate}.

    \STATE Compute $\hat R_\ell$ using
    \eqref{eq:true_range_reconstruction} and
    $\hat v_\ell$ using \eqref{eq:velocity_estimate}.
\ENDFOR

\STATE \textbf{return}
$\{(\hat R_\ell,\hat q_\ell,\hat v_\ell)\}
_{\ell=1}^{\hat L}$.

\end{algorithmic}
\end{algorithm}
\section{Performance Analysis}\label{sec:performance}

Detection, peak quality, estimation accuracy, ambiguity-order errors, and computational complexity
are analyzed below. Closed-form detection probabilities and local estimation bounds use a white Gaussian interference approximation with sampled variance $\sigma_{\mathrm s}^2$.

\subsection{Target Detection Analysis}
\label{subsec:detection}

For an isolated target, let $\mathbf t_\ell=\mathbf t_{q_\ell} (\tau_{\mathrm{app},\ell})$ and
$\mathbf g_k=[g[n-k]]_{n=0}^{N_{\mathrm{ft}}-1}$. At a fixed delay bin $k$, the transmit-reference
correlation is
\begin{equation}\label{eq:mf_peak}
z_p[k]= \mu_\ell(k)e^{-j2\pi f_{D,\ell}pT_{\mathrm{PRI}}}+u_p[k],
\qquad \mu_\ell(k)=\alpha_\ell\mathbf g_k^H\mathbf t_\ell,
\end{equation}
where $\alpha_\ell$ is the unknown complex amplitude under receive-PRI indexing. The interference samples
are independent across PRIs, with $u_p[k]\sim\mathcal{CN}(0,\sigma_u^2)$ and
$\sigma_u^2=\sigma_{\mathrm s}^2\|\mathbf g_k\|^2$. For nonzero waveforms, the normalised correlation loss is
\begin{equation}\label{eq:mismatch_loss}
L_\ell(k)=\frac{|\mathbf g_k^H\mathbf t_\ell|^2}{\|\mathbf g_k\|^2\|\mathbf t_\ell\|^2},\qquad 0\leq L_\ell(k)\leq1.
\end{equation}
The received echo energy-to-interference ratio and matched-filter output SCNR are
\begin{equation}\label{eq:scnr_pulse}
\begin{aligned}
\mathrm{SNR}_\ell&=\frac{|\alpha_\ell|^2\|\mathbf t_\ell\|^2}{\sigma_{\mathrm s}^2},\\\mathrm{SCNR}_\ell(k)
&=
\frac{|\mu_\ell(k)|^2}{\sigma_u^2}
=
L_\ell(k)\mathrm{SNR}_\ell.
\end{aligned}
\end{equation}
Correlation loss depends on waveform shape and alignment, rather than duration alone. Under the null hypothesis,
$2\mathcal T(k,f_D)/(P\sigma_u^2)$ is central chi-squared with two degrees of freedom.
The per-cell false-alarm probability is
\begin{equation}\label{eq:pfa}
P_{\mathrm{FA}} =\exp\!\left(-\frac{\eta_{\mathrm{det}}}{P\sigma_u^2}\right).
\end{equation}
At the matching Doppler frequency, its noncentrality is $2P\,\mathrm{SCNR}_\ell(k)$, giving
\begin{equation}\label{eq:pd}
P_{D,\ell}^{\mathrm{cell}}
=
Q_1\!\left(
\sqrt{2P\,\mathrm{SCNR}_\ell(k)},
\sqrt{\frac{2\eta_{\mathrm{det}}}{P\sigma_u^2}}
\right),
\end{equation}
where $Q_1$ is the first-order Marcum $Q$ function. These probabilities describe a fixed cell.
Doppler-grid mismatch reduces coherent gain; searched-grid false alarms also depend on cell dependence and peak selection. For an exactly matched reference at the correct delay, the output SCNR equals $\mathrm{SNR}_\ell$.
At the same per-cell false-alarm probability,
\begin{equation}\label{eq:pd_matched}
P_{D,\ell}^{\mathrm{matched}}= Q_1\!\left(\sqrt{2P\,\mathrm{SNR}_\ell},\sqrt{-2\ln P_{\mathrm{FA}}}\right).
\end{equation}
Each reference uses a threshold scaled to its own output interference variance. This comparison assumes that the correct propagation reference is available; initial detection uses the transmit reference.

\subsection{Peak Quality Assessment}
\label{subsec:peak_quality_perf}

PSLR and ISLR are evaluated within $\mathcal W_{\ell,p}$.
The mainlobe evaluation region $\mathcal M_\ell^{(p)}$ is centered at $k_\ell^{(p)}$, with prescribed half-width $\Delta R_m=c/(4B)$ and sample half-width $\Delta k_m=\lceil2\Delta R_m f_s/c\rceil$.
The sidelobe region is $\mathcal S_\ell^{(p)} =\mathcal W_{\ell,p}\setminus\mathcal M_\ell^{(p)}$. Using the squared magnitude adjusted by the estimated clutter and noise power in
\eqref{eq:noise_sub_energy},
\begin{equation}\label{eq:pslr_perf}
\mathrm{PSLR}_\ell^{(p)}=10\log_{10}\left(\frac{\displaystyle\max_{k\in\mathcal S_\ell^{(p)}}\tilde E_{\ell,p}[k]}
{\tilde E_{\ell,p}[k_\ell^{(p)}]}\right),
\end{equation}
\begin{equation}\label{eq:islr_perf}
\mathrm{ISLR}_\ell^{(p)}=10\log_{10}\left(\frac{\displaystyle\sum_{k\in\mathcal S_\ell^{(p)}}\tilde E_{\ell,p}[k]}
{\displaystyle\sum_{k\in\mathcal M_\ell^{(p)}}\tilde E_{\ell,p}[k]}\right).
\end{equation}
Positive denominators and nonempty sidelobe regions are required. For propagation-matched filtering,
the corresponding output and interference-power estimate are used with the same region-selection
convention. The evaluation width is a prescribed measurement convention, not a universal physical mainlobe width. Residual interference and clipping affect both metrics. Their variation with propagation distance
depends on the channel and waveform; improved sidelobe ratios alone do not establish unchanged range resolution.

\subsection{Cramér--Rao Lower Bound}
\label{subsec:crlb}
The CRLB provides a local benchmark with known ambiguity order and a fixed propagated waveform.
Let $\tilde g_\ell(t)=(g*\tilde h_\ell)(t)$ and $t_n=n/f_s$. The mean received sample is
\begin{equation}\label{eq:crlb_signal}
U_{p,n}(\boldsymbol\theta_\ell)=\alpha_\ell\tilde g_\ell(t_n-\tau_{\mathrm{app},\ell})e^{-j2\pi f_{D,\ell}pT_{\mathrm{PRI}}},
\end{equation}
where $\boldsymbol\theta_\ell=[\tau_{\mathrm{app},\ell},f_{D,\ell},\Re\{\alpha_\ell\},\Im\{\alpha_\ell\}]^T$. The channel response is held fixed when differentiating delay. The FIM is
\begin{equation}\label{eq:fim}
[\mathbf I]_{i,j}=\frac{2}{\sigma_{\mathrm s}^2}\Re\left\{\sum_{p=0}^{P-1}\sum_{n=0}^{N_{\mathrm{ft}}-1}\frac{\partial U_{p,n}^*}{\partial\theta_i}\frac{\partial U_{p,n}}{\partial\theta_j}\right\}.
\end{equation}

After eliminating the unknown complex amplitude,the equivalent delay information is
\begin{equation}\label{eq:delay_information}
I_{\tau\tau}^{\mathrm{eff}}=\frac{2P|\alpha_\ell|^2}{\sigma_{\mathrm s}^2}
\left(\|\dot{\mathbf t}_\ell\|^2-\frac{|\mathbf t_\ell^H\dot{\mathbf t}_\ell|^2}{\|\mathbf t_\ell\|^2}
\right),
\end{equation}
where
$\dot{\mathbf t}_\ell=[-\tilde g_\ell'(t_n-\tau_{\mathrm{app},\ell})]_n$.
For the separable fast-time and slow-time model, the equivalent cross term is zero, and
\begin{equation}\label{eq:fim_diag}
I_{ff}^{\mathrm{eff}}=8\pi^2T_{\mathrm{PRI}}^2\mathrm{SNR}_\ell\frac{P(P^2-1)}{12}.
\end{equation}
Eliminating the common unknown phase centers the slow-time indices, yielding the factor $P(P^2-1)/12$ consistent with the classical single-tone bound \cite{1055282}. For a sufficiently captured band-limited waveform, let $W_\ell(\nu)=|G(\nu)|^2 |\tilde H_\ell(f_c+\nu)|^2$. Its effective RMS bandwidth is
\begin{equation}\label{eq:alpha_eff}
\begin{aligned}
\bar\nu_\ell&=\frac{\int\nu W_\ell(\nu)\,\mathrm d\nu}{\int W_\ell(\nu)\,\mathrm d\nu},\\
\alpha_{\mathrm{eff},\ell}^2&=
\frac{\int(\nu-\bar\nu_\ell)^2
W_\ell(\nu)\,\mathrm d\nu}
{\int W_\ell(\nu)\,\mathrm d\nu}.
\end{aligned}
\end{equation}
The range and velocity bounds are
\begin{equation}\label{eq:crlb_range}
\mathrm{CRLB}(R_\ell)=
\frac{c^2}{4I_{\tau\tau}^{\mathrm{eff}}}
\simeq
\frac{c^2}
{8(2\pi\alpha_{\mathrm{eff},\ell})^2
\mathrm{SNR}_\ell P},
\end{equation}
\begin{equation}\label{eq:crlb_velocity}
\mathrm{CRLB}(v_\ell)
=\frac{3c^2}{8\pi^2f_c^2T_{\mathrm{PRI}}^2\mathrm{SNR}_\ell P(P^2-1)}.
\end{equation}

For a sinc spectrum truncated to $|\nu|\leq B/2$ with $BT_p=1$ and a constant in-band transmit-filter
response, $\alpha_0\simeq0.256B$. Using $\alpha_{\mathrm{eff},\ell}\simeq\alpha_0$ requires negligible change in normalised spectral magnitude. Spectral phase alone does not change this bandwidth; frequency-selective attenuation can. These bounds apply to locally unbiased estimation under the fixed-waveform approximation. They exclude order-selection errors and do not guarantee attainment by the sequential receiver. The velocity bound describes variance, whereas the conventional Doppler-bin spacing corresponds
to $\Delta v=\lambda/(2PT_{\mathrm{PRI}})$.

\subsection{Probability of Error}
\label{subsec:perr}

For nonzero aligned references, their similarity is
\begin{equation}\label{eq:rho_adj}
\rho_{q,q'}(\tau)
=
\frac{|\mathbf t_q^H(\tau)\mathbf t_{q'}(\tau)|}
{\|\mathbf t_q(\tau)\|\|\mathbf t_{q'}(\tau)\|}.
\end{equation}
The fraction $1-\rho_{q,q'}^2$ is the normalised energy orthogonal to the competing reference.
When $\rho_{q,q'}=1$, the unknown complex amplitude makes the two references indistinguishable
at that delay. Let $A_\ell(q)=\max_{f_D\in\mathcal F_D}S_\ell(q;\hat\tau_{\mathrm{app},\ell}^{\mathrm{init}},f_D)$. For the decision in \eqref{eq:q_hat_accum},
\begin{equation}\label{eq:pairwise}
P_{q\rightarrow q'}
=
\Pr\!\left[
A_\ell(q')\geq A_\ell(q)
\,\middle|\,q_\ell=q
\right],
\end{equation}
and
\begin{equation}\label{eq:order_error_union}
P_{\mathrm{err}\mid q}
\leq
\sum_{\substack{q'=0\\q'\neq q}}^{q_{\max}}
P_{q\rightarrow q'}.
\end{equation}
These probabilities depend on interference, initial-delay error, Doppler search, and channel-model accuracy. Reference similarity alone does not determine a universal identification-SCNR penalty. For valid measurements in the scalar interval classifier, assume
$\hat\beta_\ell^{(p)}
\approx\mathcal N(\mu_{\beta,\ell},
\sigma_{\hat\beta,\ell}^2)$.
The approximate conditional probability of correct classification is
\begin{equation}\label{eq:p0}
p_{0,\ell}
=
\Phi\!\left(
\frac{b_{q_\ell+1}-\mu_{\beta,\ell}}
{\sigma_{\hat\beta,\ell}}
\right)
-
\Phi\!\left(
\frac{b_{q_\ell}-\mu_{\beta,\ell}}
{\sigma_{\hat\beta,\ell}}
\right),
\end{equation}
where $\Phi$ is the standard normal CDF. The mean permits measurement bias; unconditional performance also includes undecided measurements. This scalar approximation does not describe the main correlation-based decision.

\subsection{Computational Complexity}\label{subsec:complexity}
For a uniform $P$-point Doppler grid, conventional FFT-based processing costs $O(PN_{\mathrm{ft}}\log N_{\mathrm{ft}} +N_{\mathrm{ft}}P\log P)$ per CPI. Order identification adds $O(\hat L(q_{\max}+1)PN_{\mathrm{ft}})$ for inner products and $O(\hat L(q_{\max}+1)P\log P)$ for slow-time transforms. Online inverse-FFT template synthesis additionally costs $O(\hat L(q_{\max}+1) N_{\mathrm{ft}}\log N_{\mathrm{ft}})$. With $N_\tau$ directly evaluated delay trials per detection, selected-reference estimation costs $O(\hat LPN_\tau N_{\mathrm{ft}})$, followed by $O(\hat LP\log P)$ for Doppler. FFT-based delay correlation instead costs $O(\hat LPN_{\mathrm{ft}}\log N_{\mathrm{ft}})$.
A uniform zero-padded Doppler grid of length $N_D\geq P$ replaces each slow-time FFT cost by
$O(N_D\log N_D)$. The scalar classifier substitutes duration measurement and threshold lookup for candidate
correlations. Comparisons with diversity methods must include their waveform banks, pulse allocations,
sampling schedules, and order-reconstruction costs. Relative complexity therefore depends on
$\hat L$, $q_{\max}$, search sizes, and implementation.
\section{Simulation Results and Discussions}\label{sec:results}
In this section, the proposed framework is evaluated against the configuration of Table~\ref{tab:params}. Two of the five targets lie beyond $R_u$ and fold to apparent ranges of $3.6$ and $6.9$~m. 
\begin{table}[t]
\centering
\caption{Simulation Parameters}
\label{tab:params}
\renewcommand{\arraystretch}{1.15}
\setlength{\tabcolsep}{4pt}
\begin{tabular}{|l|l|}
\hline
\textbf{Parameter} & \textbf{Value} \\
\hline
Carrier frequency
& $f_c = 470$~GHz \\
Signal bandwidth
& $B = 94$~GHz \\
Nominal pulse duration
& $T_p = 10.64$~ps \\
Pulse repetition interval
& $T_{\mathrm{PRI}} = 66.71$~ns \\
Unambiguous range
& $R_u \approx 10$~m \\
Maximum modeled range
& $R_{\max} = 30$~m \\
Maximum ambiguity order
& $q_{\max} = 2$ \\
Sampling rate
& $f_s = 188$~GS/s \\
Pulses per CPI
& $P = 1024$ \\
Ambient temperature
& $T = 296$~K \\
Atmospheric pressure
& $p_{\mathrm{atm}} = 101.325$~kPa \\
Water-vapour density
& $\rho_w = 10$~g/m$^3$ \\
True target ranges
& $\{2.4,\,5.2,\,8.1,\,13.6,\,26.9\}$~m \\
Radial velocities
& $\{0.5,\,-1.2,\,2.0,\,-3.1,\,4.4\}$~m/s \\
Declared relative target levels
& $\{0,\,-2,\,-4,\,-6,\,-8\}$~dB \\
Per-cell false-alarm probability
& $P_{\mathrm{FA}} = 10^{-7}$ \\
Clutter scatterer density
& $\lambda_c = 0.8$~m$^{-1}$ \\
Monte Carlo trials
& $N_{\mathrm{MC}} = 2000$ \\
\hline
\end{tabular}
\end{table}
\begin{figure*}[ht]
\centering
\subfloat[]{ \includegraphics[width=55mm,height=40mm]{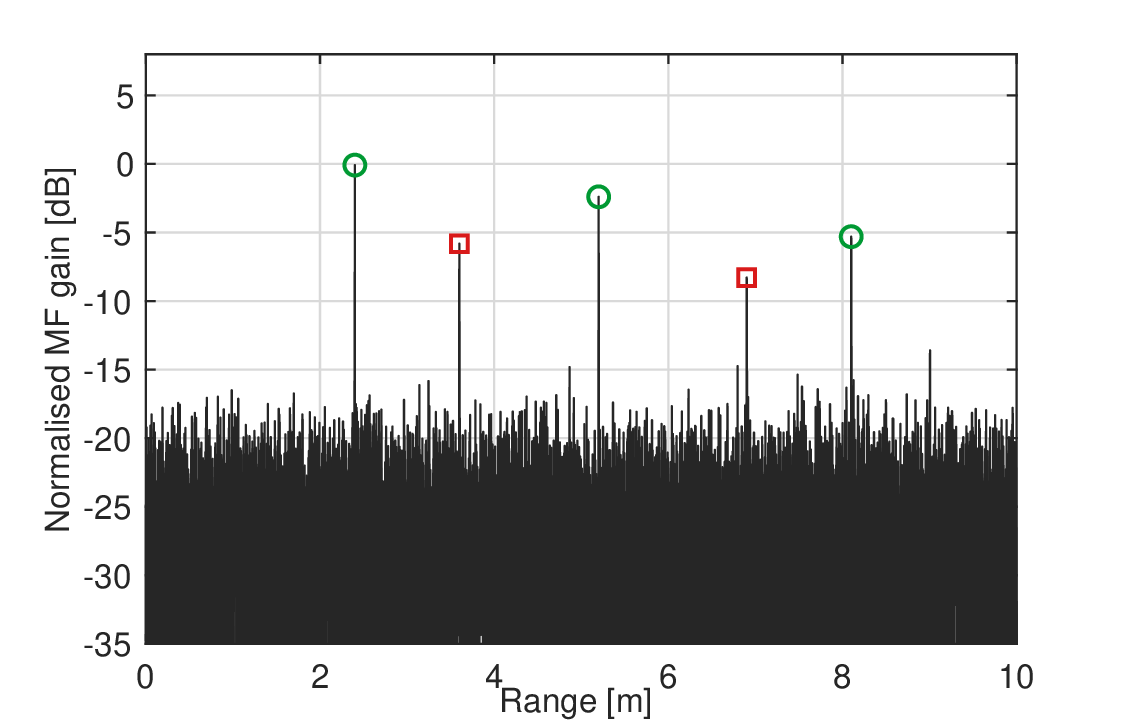}}%
\subfloat[]{ \includegraphics[width=55mm,height=40mm]{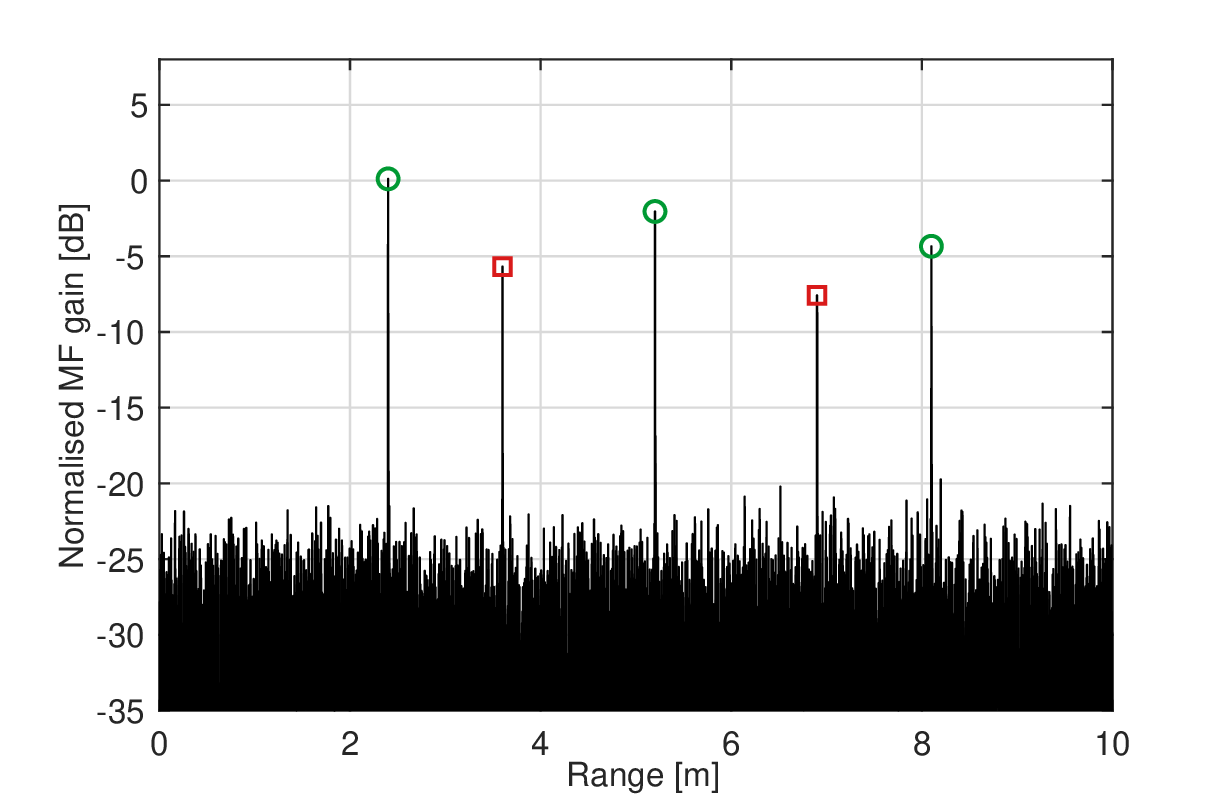}}%
\subfloat[]{ \includegraphics[width=55mm,height=40mm]{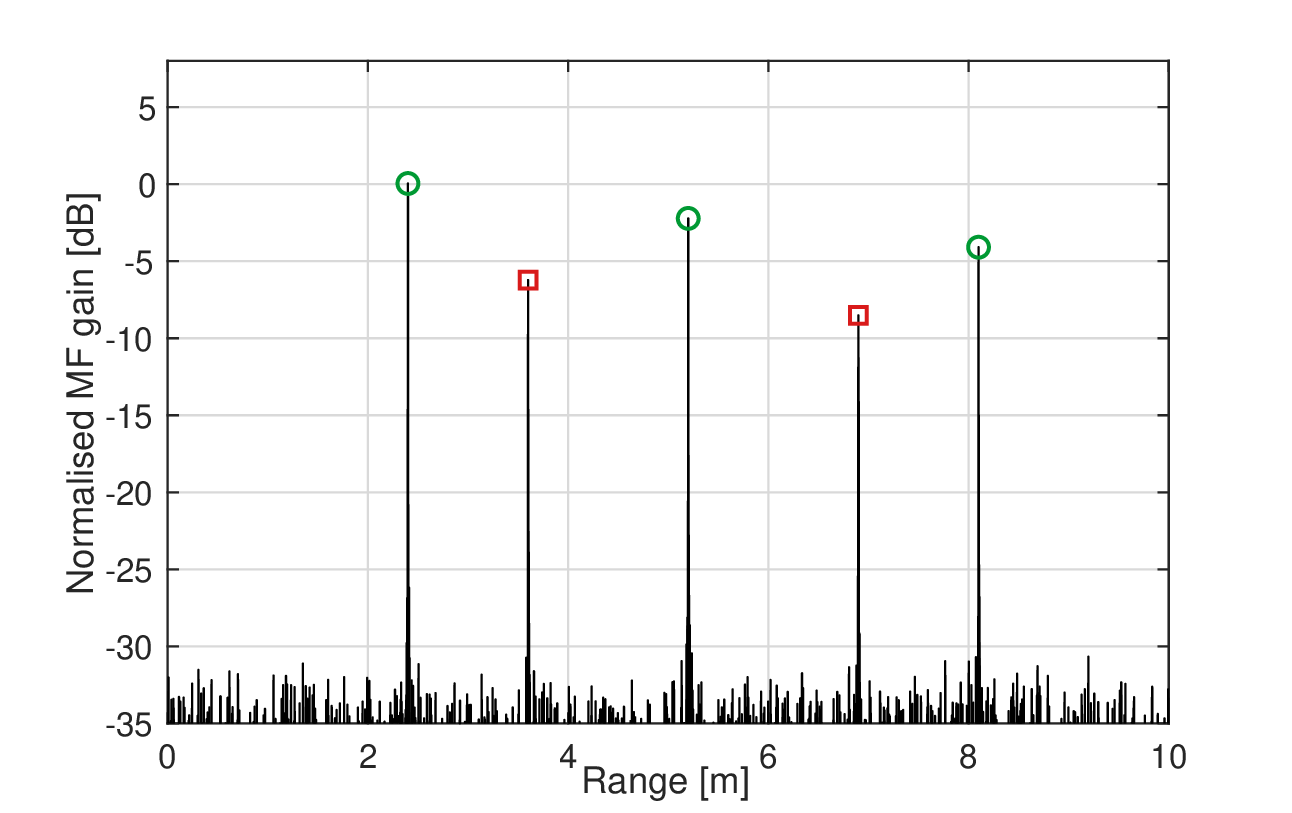}}\\

\subfloat[]{ \includegraphics[width=55mm,height=40mm]{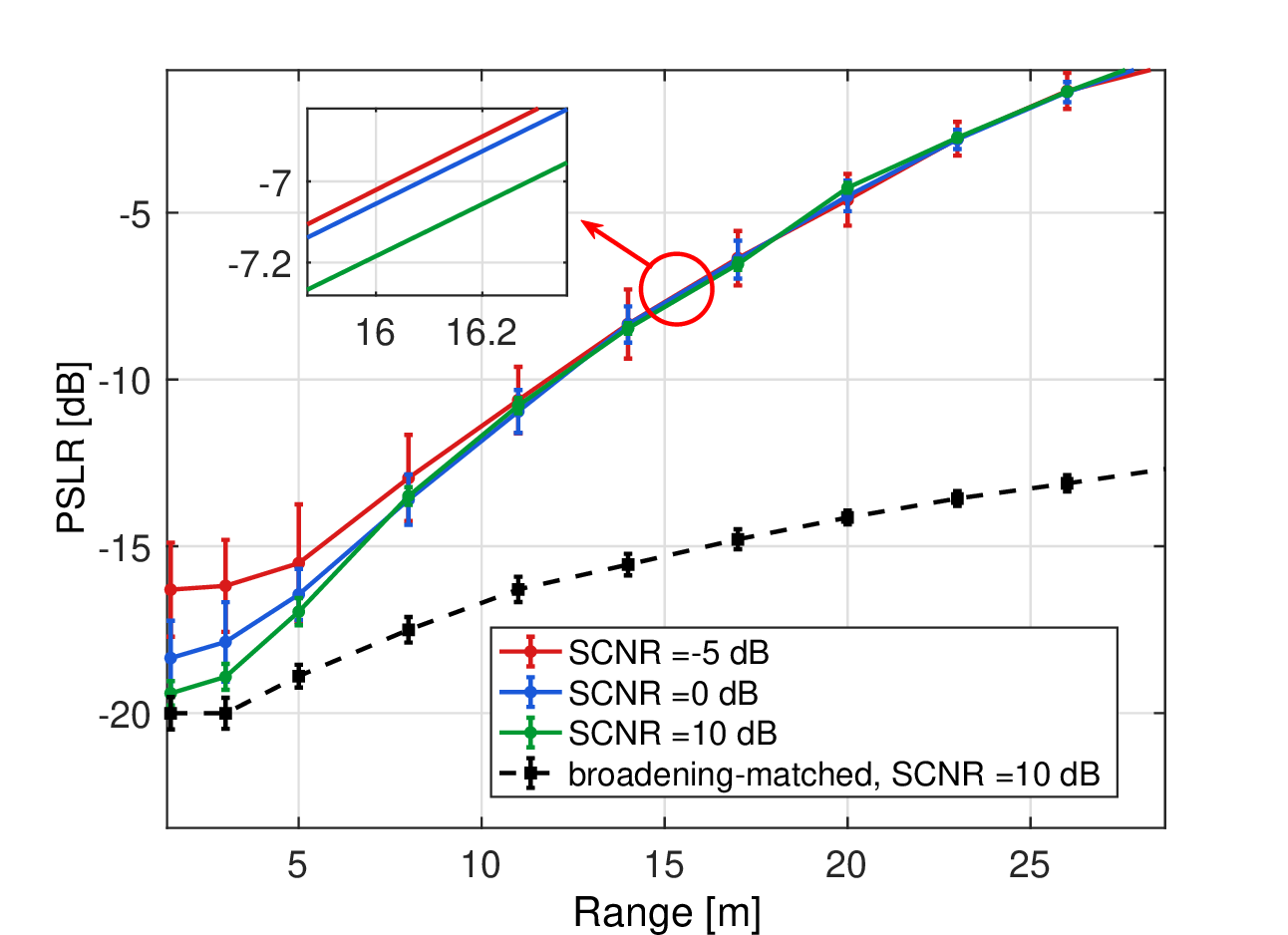}}
\subfloat[]{ \includegraphics[width=55mm,height=40mm]{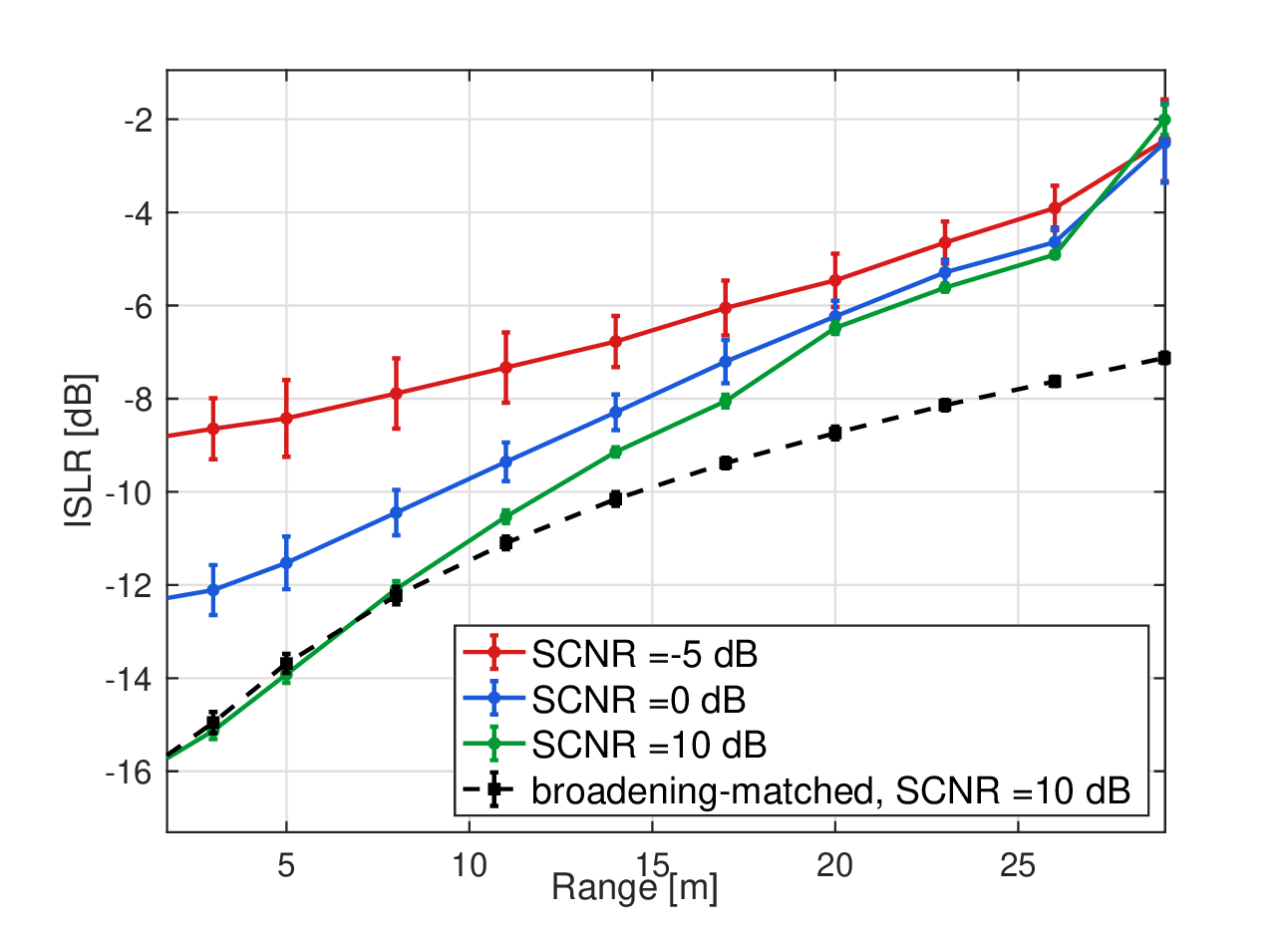}}
\subfloat[]{ \includegraphics[width=55mm,height=40mm]{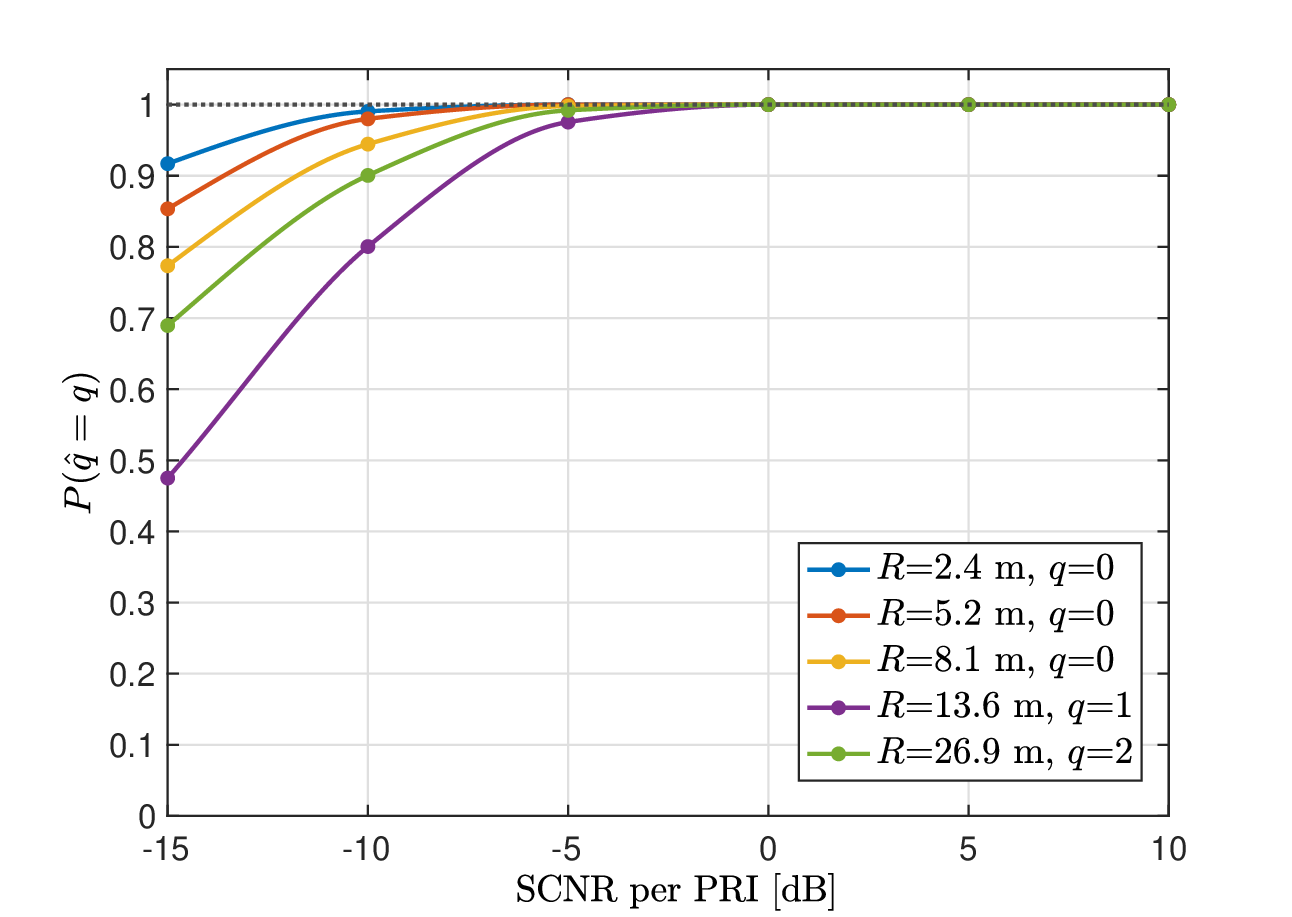}}
\caption{Range profile of the normalised matched-filter gain over one PRI, comprising three unambiguous targets (green circles) and two folded targets (red squares), at per-PRI SCNR of (a) $-5$~dB, (b) $0$~dB and (c) $10$~dB; (d) PSLR vs range, (e) ISLR vs range, with the broadening-matched reference, and (f) Probability of correct ambiguity order identification vs SCNR per PRI.}
\label{fig:profile}
\end{figure*}
\subsection{Range Profile Analysis}
Fig.~2(a)--(c) presents the normalised matched-filter range profiles over one unambiguous-range interval at per-PRI SCNR values of $-5$, $0$, and $10$~dB. The delay axis spans one PRI, whereas initial detection combines the PRI-wise correlation outputs coherently across the CPI with Doppler compensation. With $R_u \approx 10$~m, the targets at $2.4$, $5.2$, and $8.1$~m lie within the unambiguous interval and have ambiguity order $q_\ell=0$. The targets at $13.6$ and $26.9$~m have orders $q_\ell=1$ and $q_\ell=2$, respectively, and appear at
approximately $3.6$ and $6.9$~m. Their echoes therefore occupy distinct apparent-delay locations interleaved with those of the unambiguous targets. Delay-only processing associates these returns with their apparent ranges and cannot recover the corresponding physical ranges. As SCNR increases, the background fluctuations decrease
relative to the target peaks, making the apparent-delay locations more clearly distinguishable. The folded locations
persist at all three operating points because the PRI determines the folding relationship. The folded returns
also exhibit lower normalised peak levels in the illustrated profiles. These levels depend on the received target
strengths and the mismatch between the propagated echoes and the transmit reference, and therefore do not uniquely
identify the ambiguity orders. Following detection, the proposed receiver compares propagation-matched references for the candidate physical ranges using the normalised, Doppler-compensated coherent correlation statistic. The selected order is then retained during matched-reference delay refinement and physical-range reconstruction. Fig.~2 (f) evaluates the reliability of this order-selection stage statistically.

\subsection{PSLR and ISLR Versus True Range}
\label{subsec:pslr_results}
Fig.~\ref{fig:profile}(d)--(e) presents PSLR and ISLR versus true target range for the transmit reference at per-PRI SCNR values of $-5$, $0$, and $10$~dB, together with the broadening-matched reference at $10$~dB. Both metrics use the prescribed mainlobe and sidelobe regions of Section~V-B. At $10$~dB, the transmit-reference PSLR increases from $-19.43$~dB at $1.5$~m to $-0.14$~dB at $29$~m, while ISLR increases from $-15.76$ to $-1.98$~dB. This degradation is consistent with increasing mismatch between the distance-dependent propagated echo and the fixed transmit reference, which reduces the concentration of correlation energy within the prescribed mainlobe region. At $26.9$~m, the PSLR reaches $-0.94$~dB, indicating that the largest sidelobe is comparable in power to the mainlobe. Returns sharing an apparent delay can consequently exhibit different peak qualities: at an apparent range of $3.6$~m, the returns from true ranges of $3.6$ and $13.6$~m yield PSLR values of $-18.48$ and $-8.78$~dB, respectively. The PSLR curves exhibit similar range dependence across the three SCNR values, whereas ISLR is more sensitive to SCNR at short range. At $1.5$~m, ISLR is $-8.87$~dB at $-5$~dB SCNR and $-15.76$~dB at $10$~dB SCNR. This difference is consistent with the sensitivity of the summed sidelobe energy to residual interference after power subtraction and nonnegative clipping. Beyond approximately $20$~m, the ISLR curves become closer as the
distance-dependent sidelobe contribution becomes more prominent. At $10$~dB SCNR, propagation-matched filtering improves PSLR and ISLR across the evaluated range, with improvements of $12.7$ and $5.2$~dB, respectively, at $29$~m. The matched-reference PSLR nevertheless varies from $-20.31$ to $-12.79$~dB. Matching reduces reference mismatch, while the received spectral magnitude and the prescribed evaluation regions continue to determine the sidelobe metrics. This comparison assumes correct ambiguity-order selection and quantifies the peak-quality improvement available during subsequent matched-reference estimation.

\begin{figure*}[ht]
\centering\subfloat[]{ \includegraphics[width=50mm,height=40mm]{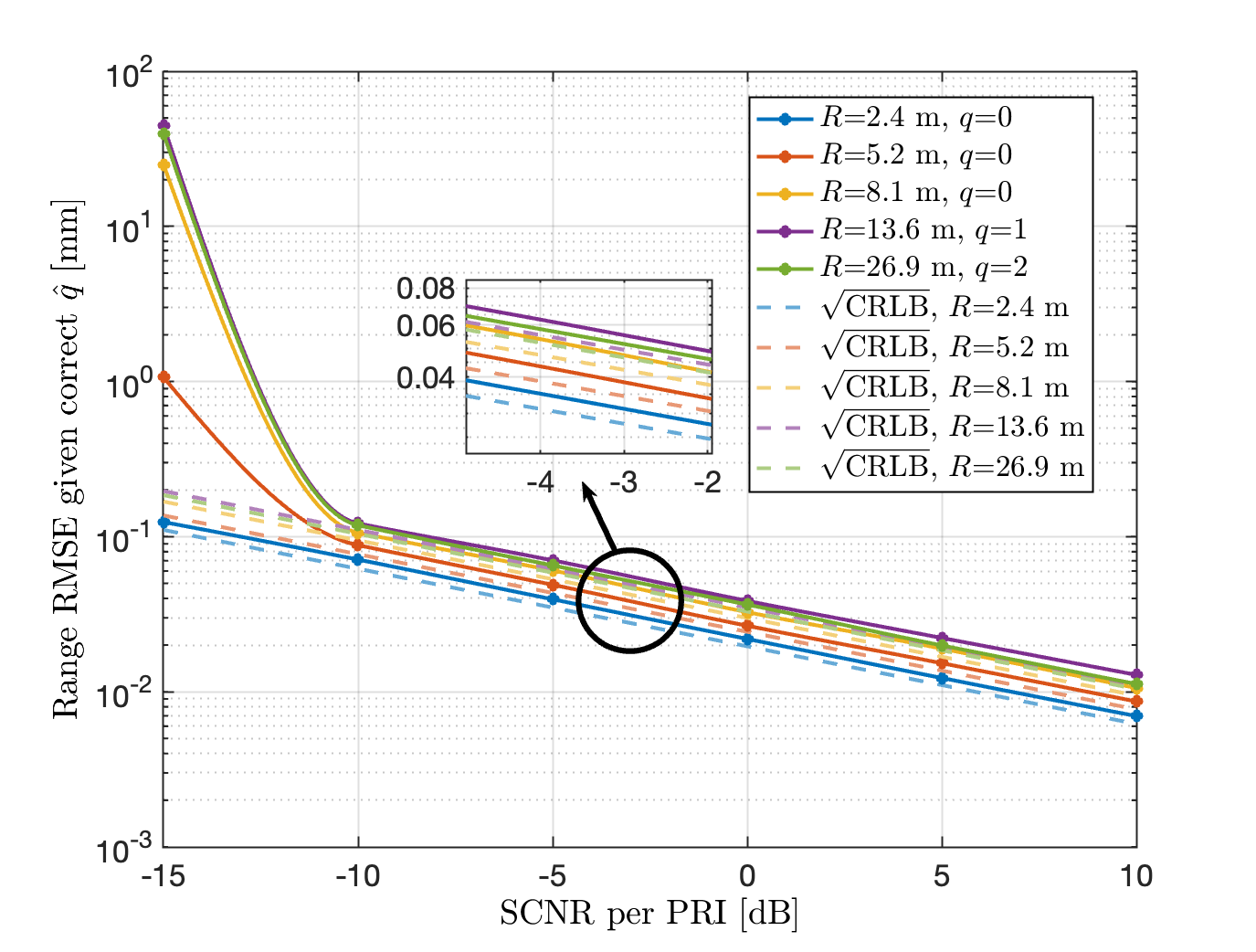}}%
\subfloat[]{ \includegraphics[width=55mm,height=40mm]{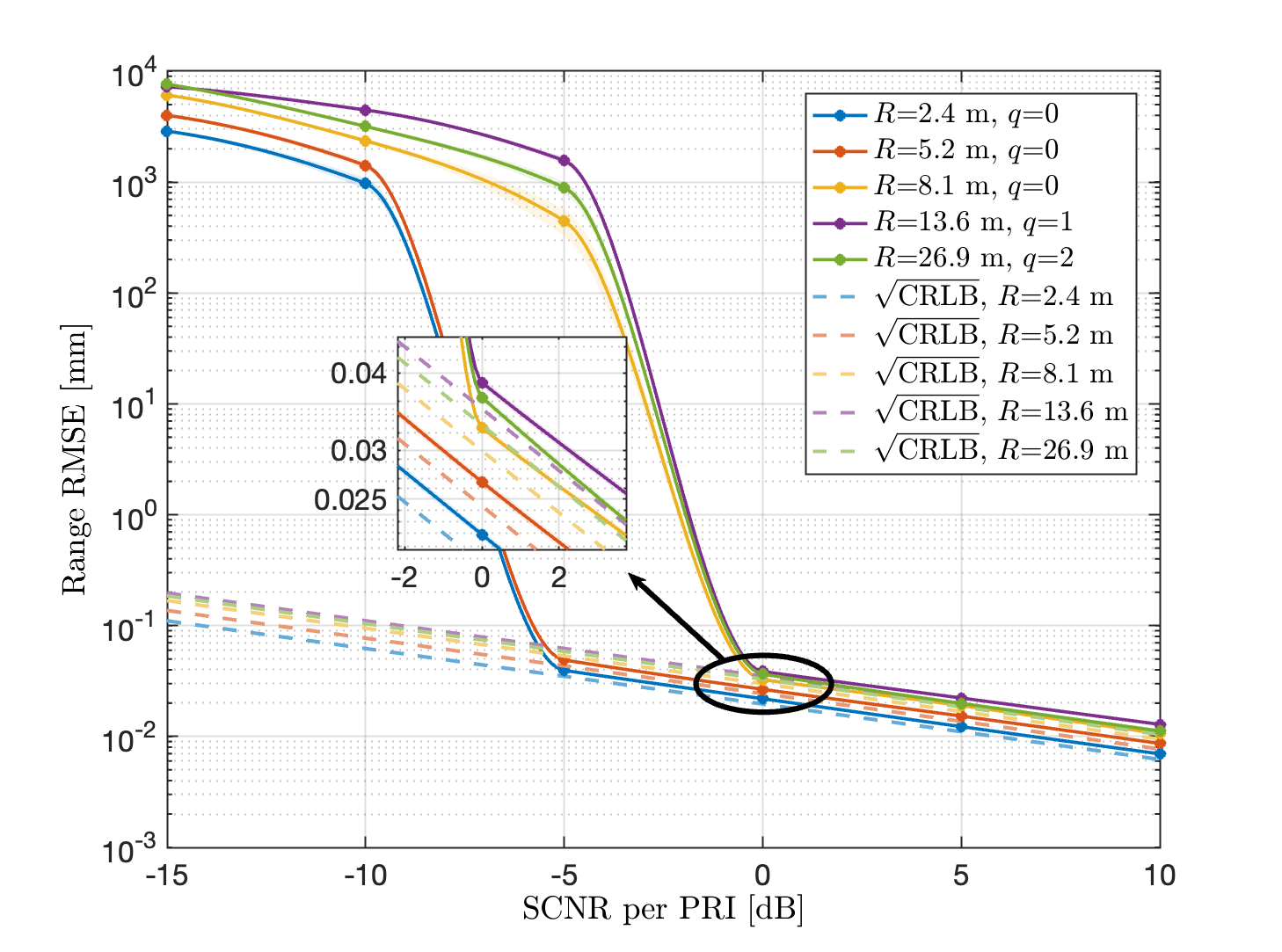}}%
 \subfloat[]{ \includegraphics[width=55mm,height=40mm]{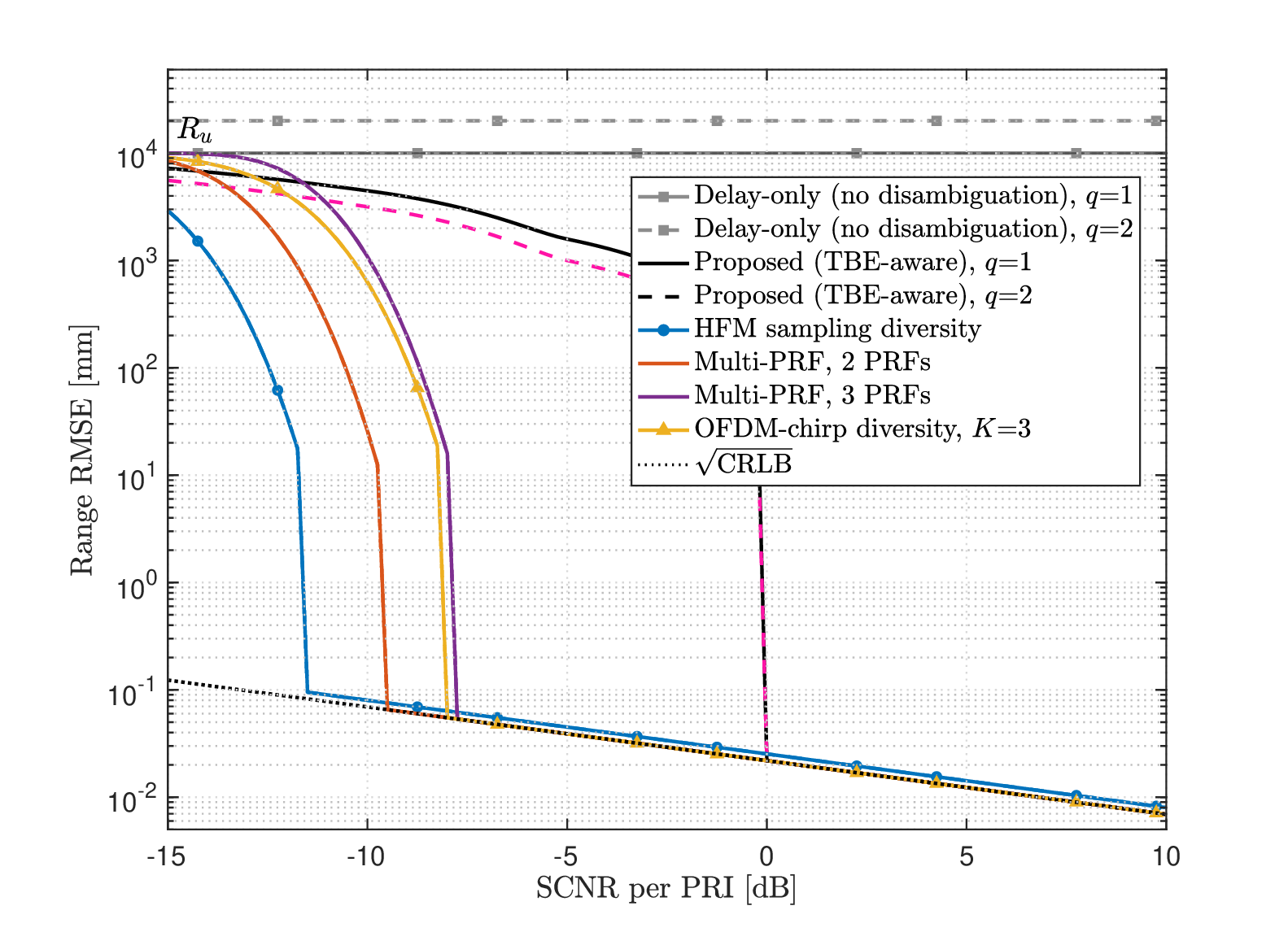}}\\
 \subfloat[]{ \includegraphics[width=55mm,height=40mm]{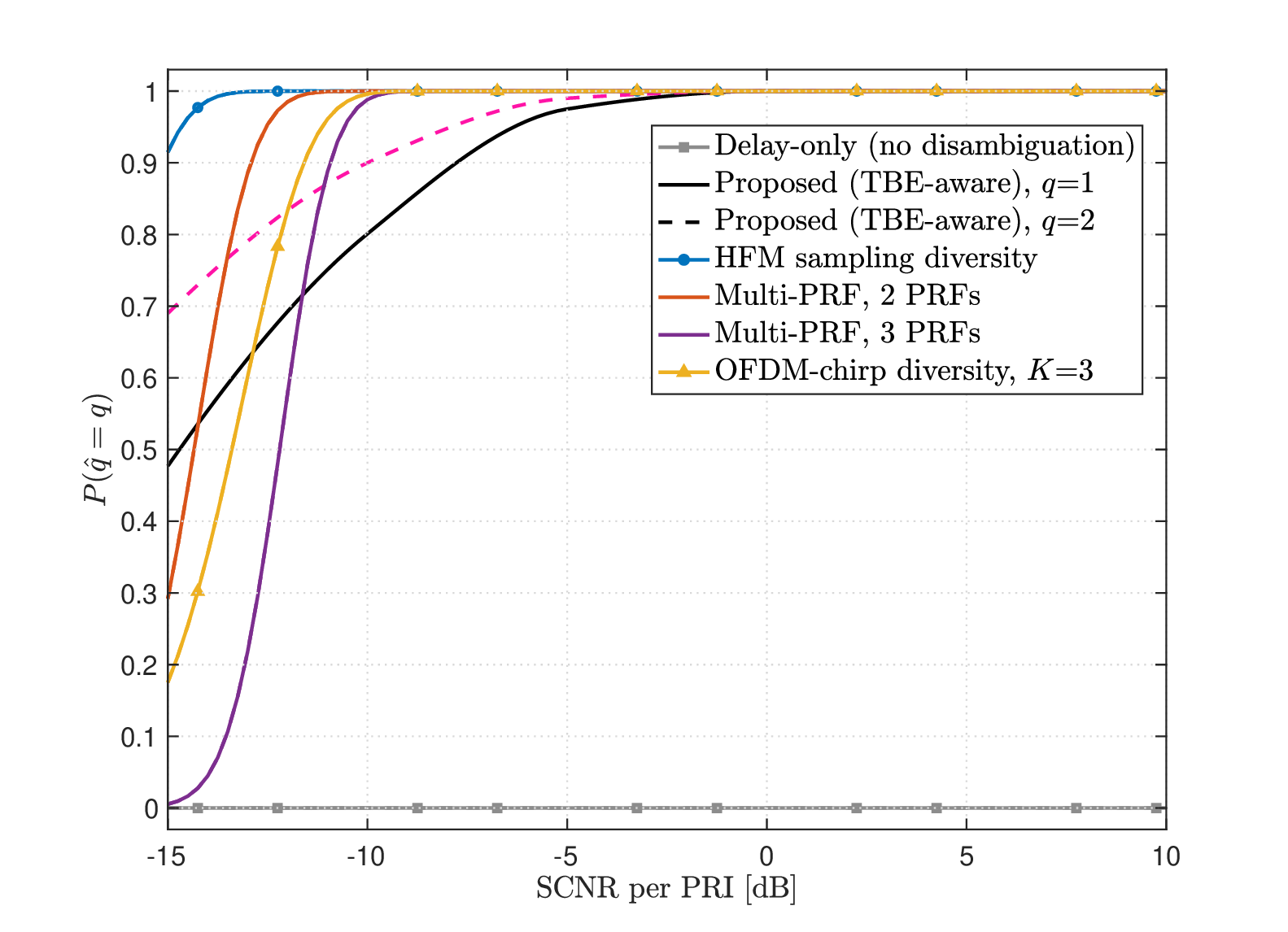}}
\subfloat[]{ \includegraphics[width=55mm,height=40mm]{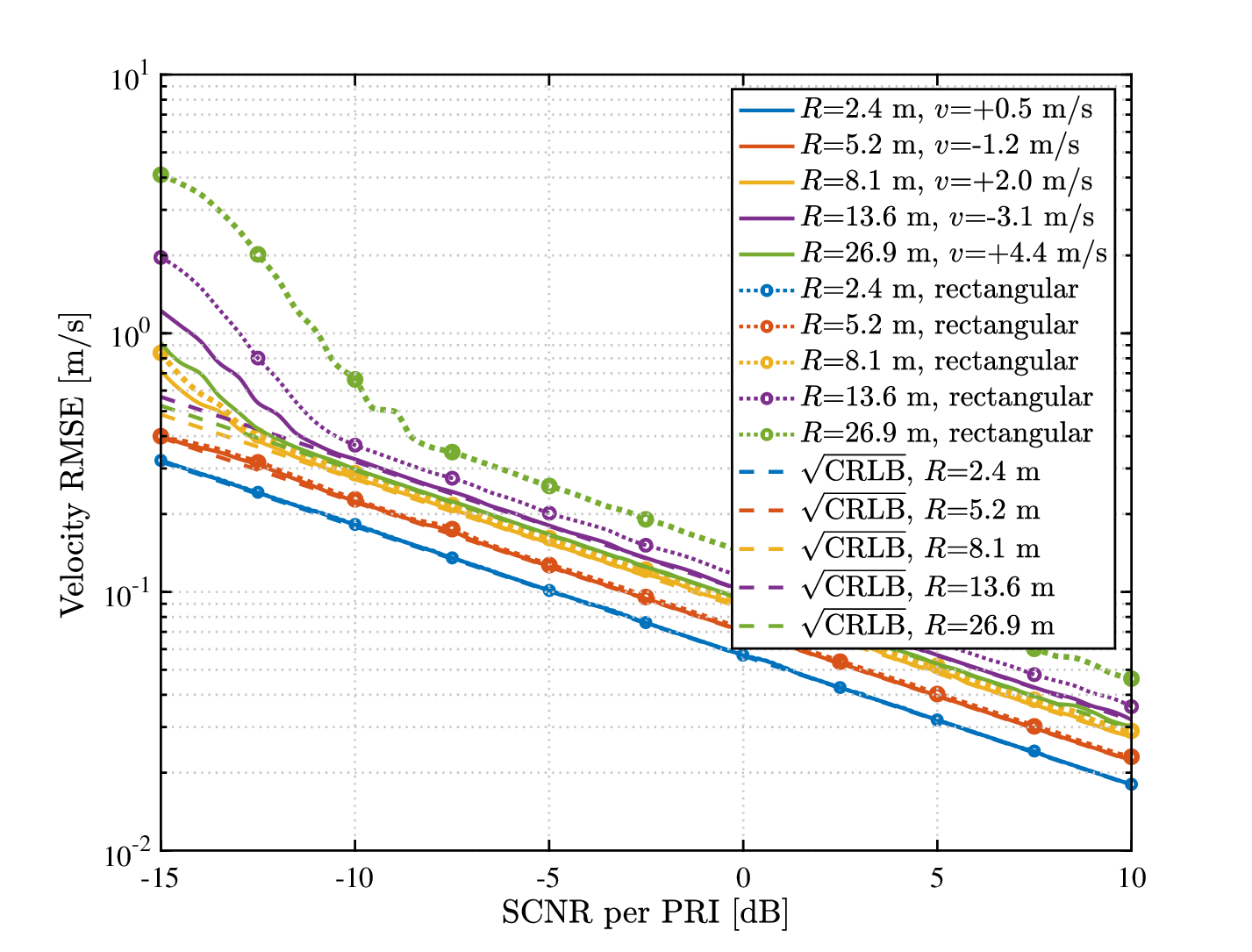}}
\subfloat[]{ \includegraphics[width=55mm,height=40mm]{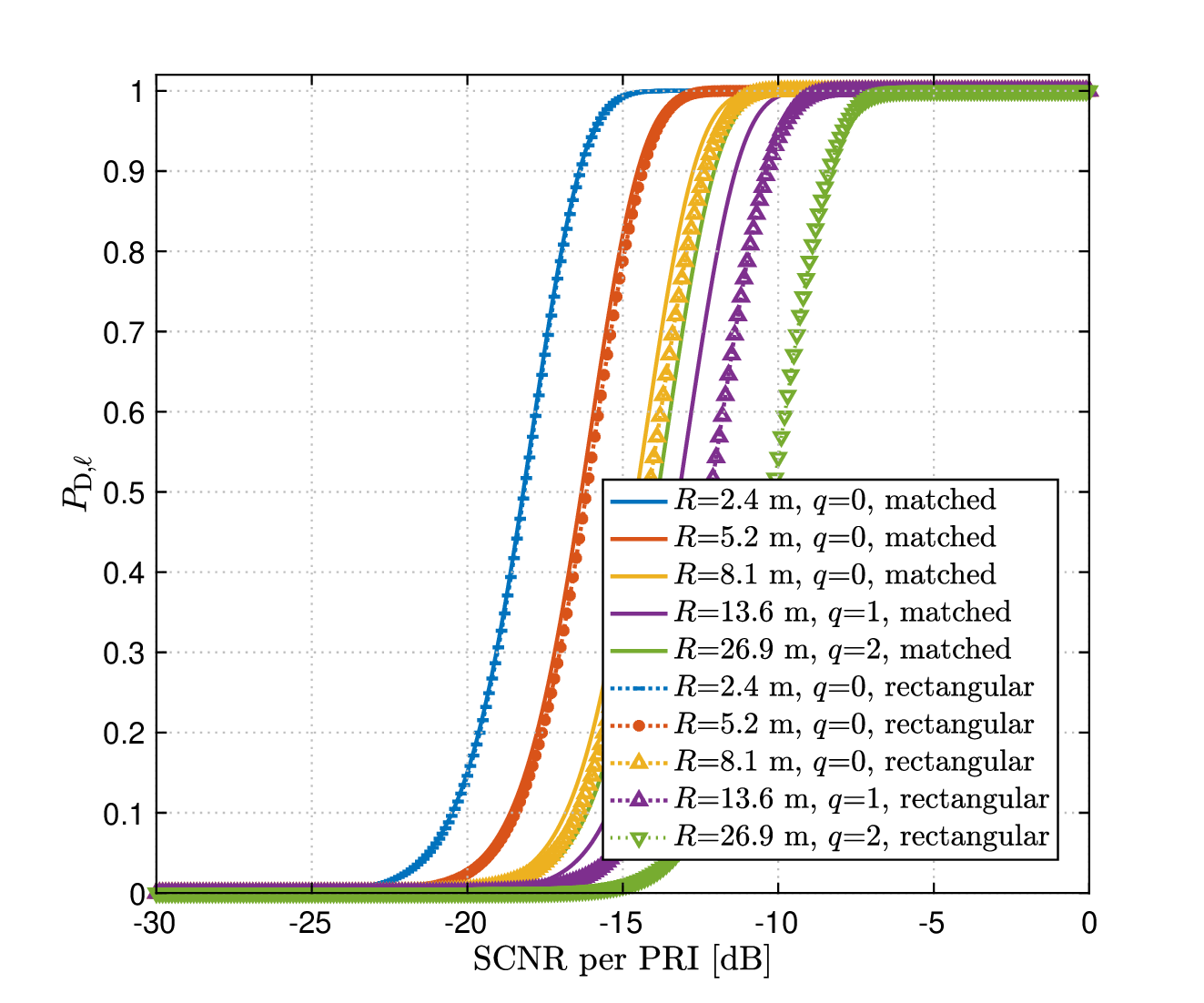}}
\caption{(a) range RMSE of the proposed TBE-aware receiver conditioned
on correct ambiguity-order identification, together with the
corresponding CRLB; (b) unconditional range RMSE of the proposed
receiver; (c) unconditional range RMSE comparison with prior
ambiguity-resolution methods; (d) probability of correct
ambiguity-order identification versus per-PRI SCNR for the proposed and
benchmark schemes; (e) velocity RMSE for the rectangular and
broadening-matched references, with the corresponding CRLB; and (f)
probability of detection for both references at
$P_{\mathrm{FA}}=10^{-7}$.} \label{fig:rmse}
\end{figure*}

\subsection{Ambiguity-Order Identification}\label{subsec:identification}
Fig.~\ref{fig:profile}(f) presents the empirical probability of correct ambiguity-order identification using \eqref{eq:q_hat_accum}, with all $P=1024$ received PRIs combined coherently with Doppler compensation. Identification improves with SCNR for all five targets. At $-15$~dB, the probabilities for the unambiguous targets at $2.4$, $5.2$, and $8.1$~m are $0.915$, $0.855$, and $0.778$, respectively. The folded targets at $13.6$ and $26.9$~m attain $0.477$ and $0.690$. The target at $13.6$~m remains the most difficult to identify, with an error rate of approximately $2.5\%$ at $-5$~dB. This ordering is consistent with the dependence of identification reliability on candidate-reference similarity discussed in Section~\ref{subsec:perr}. For the target at $13.6$~m, the apparent range is approximately $3.6$~m, giving candidate physical ranges of $3.6$, $13.6$, and $23.6$~m. The correct hypothesis can therefore be confused with candidates on either side of its physical range. The corresponding error probabilities depend on reference similarity, effective target SCNR, and initial-delay alignment. Normalising the coherent correlations removes arbitrary scalar scaling of the candidate references, while their discrimination remains sensitive to interference. Increasing SCNR and coherent integration improve the reliability of the comparison. At SCNR values of $0$~dB and above, all curves attain an empirical probability of one, indicating no observed identification errors in the reported trials. With $N_{\mathrm{MC}}=2000$ independent trials per target, zero observed errors corresponds to a one-sided $95\%$ upper confidence limit of approximately $3/N_{\mathrm{MC}}=1.5\times10^{-3}$ on the true error
probability. These results apply to targets occupying distinct apparent-delay bins. If returns from different ambiguity orders overlap in the same bin, their coherent superposition does not generally correspond to a single candidate reference. The decision in \eqref{eq:q_hat_accum} selects one order and cannot separately recover both components. Resolving such returns requires a multi-component estimator and sufficiently distinguishable propagation signatures.

\subsection{Range Accuracy Analysis}\label{subsec:range_accuracy}
Fig.~\ref{fig:rmse}(a) presents the range RMSE conditioned on correct ambiguity-order identification, together with
$\sqrt{\mathrm{CRLB}(R_\ell)}$ from \eqref{eq:crlb_range}, evaluated using the received echo-energy ratio $\mathrm{SNR}_\ell$. This local benchmark assumes a known ambiguity order and a fixed propagated waveform. Conditioning excludes order-selection errors from the reported RMSE, while delay-estimation errors remain. At low SCNR, the measured errors exceed the bounds, consistent with threshold behaviour in delay estimation, including noise-driven peak selection and residual reference-alignment mismatch. As SCNR increases, the curves approach their corresponding bounds and exhibit approximately inverse-square-root dependence on the linear SCNR, reaching $0.022$--$0.038$~mm at $0$~dB. The differences between the target-specific bounds depend on both received echo energy and effective RMS bandwidth. These results demonstrate the local estimation accuracy achievable after successful identification under the assumed propagation model. 
Fig.~\ref{fig:rmse}(b) includes ambiguity-order errors. Through \eqref{eq:true_range_reconstruction}, an incorrect order introduces an additional range-error term equal to $(\hat q_\ell-q_\ell)R_u$. Consequently, even infrequent order errors can dominate the range MSE. At $-15$~dB, the unconditional RMSE is approximately $3$--$8$~m and decreases rapidly as identification becomes more reliable. The target at $13.6$~m exhibits the latest transition to small errors, consistent with its lower identification probability in Fig.~\ref{fig:profile}(f). At $0$~dB and above, the conditional and unconditional curves become closely aligned in the reported trials. The transition to accurate physical-range reconstruction is therefore governed primarily by ambiguity-order identification in this scene.
Fig.~\ref{fig:rmse}(c) compares the unconditional range RMSE with the benchmark schemes. Delay-only processing retains range biases of approximately $R_u$ and $2R_u$ for the two folded targets. The ambiguity-aware schemes enter the local delay-estimation regime as order-selection errors become rare. Using the nominal resolution cell $c/(2B)=1.59$~mm, the empirical crossing points are $-11.52$~dB for the HFM sampling-diversity benchmark, $-9.53$~dB for two PRFs, $-8.02$~dB for OFDM-chirp,
and $-7.77$~dB for three PRFs. The proposed receiver falls below this level by $0$~dB in the reported trials.
At higher SCNR, the curves become closely spaced, indicating comparable local ranging accuracy despite different identification requirements. These crossing points require a finite-sample qualification. For predominantly single-fold mistakes and local delay errors much smaller than $R_u$, the ambiguity-error contribution to MSE is approximately
$R_u^2\Pr(\hat q_\ell\neq q_\ell)$. With $R_u\approx10$~m, a population RMSE below $1.59$~mm consequently requires
an order-error probability of approximately $2.5\times10^{-8}$ or less. Such probabilities cannot
be established from $N_{\mathrm{MC}}=2000$ ordinary Monte Carlo trials. The reported sub-cell crossings
therefore describe empirical performance; they do not establish population-level thresholds for any of the compared schemes.

\subsection{Identification Against Prior Methods} \label{subsec:identification_compare}
Fig.~\ref{fig:rmse}(d) compares ambiguity-order identification with numerical reference benchmarks
comprising two- and three-PRF processing, OFDM-chirp diversity with $K=3$, and HFM sampling diversity. These techniques are applied in the considered THz scenario, while the proposed receiver uses the fixed transmission and sampling configuration of Table~\ref{tab:params}. The reported thresholds characterise the evaluated implementations and do not establish a performance ranking under identical resource constraints. Delay-only processing assigns the apparent range without identifying the folding order and therefore fails to identify either folded target. For the target with $q_\ell=1$, the proposed receiver reaches an identification probability of $0.99$ at $-3.02$~dB per-PRI SCNR. The corresponding reported thresholds
are $-9.93$~dB for three PRFs, $-10.34$~dB for the $K=3$ OFDM-chirp scheme, $-11.83$~dB for two PRFs, and $-13.86$~dB for the HFM reference benchmark. The proposed receiver thus requires higher SCNR for this operating point while retaining a fixed
transmitted pulse, PRF, and sampling rate. This higher identification requirement is consistent with the similarity between propagation-matched candidate references. For a competing reference pair with $\rho_{q,q'}=0.945$, the normalised energy fraction orthogonal to the competing reference is $1-\rho_{q,q'}^2\approx0.107$. This limited separation
makes the correlation comparison sensitive to interference and alignment errors, as discussed in Section~\ref{subsec:perr}. Reference similarity characterises the discrimination difficulty, while the pairwise error probabilities in \eqref{eq:pairwise} also depend on the adopted interference and search conditions. The comparison illustrates the identification-SCNR cost of using a propagation-induced signature. Multi-PRF and OFDM-chirp processing introduce diversity through transmission, whereas the HFM benchmark uses receiver-side sampling diversity. The proposed receiver instead distinguishes candidate physical ranges through their propagation responses and combines the correlations coherently across the CPI. The numerical ranking remains specific to the waveform, sampling, propagation, and resource assumptions of the evaluated configurations.

\subsection{Velocity Accuracy}\label{subsec:velocity_accuracy}
Fig.~\ref{fig:rmse}(e) compares the velocity RMSE for the transmit and propagation-matched references with
$\sqrt{\mathrm{CRLB}(v_\ell)}$. At $0$~dB, the matched-reference velocity RMSE is $0.057$--$0.099$~m/s, close to the corresponding bounds of $0.057$--$0.101$~m/s. These values quantify local estimation accuracy, whereas the conventional Doppler-bin spacing $\Delta v=\lambda/(2PT_{\mathrm{PRI}})=4.67$~m/s is set by the CPI duration. For fixed reference alignment and a stable slow-time response, the transmit-reference mismatch reduces the output SCNR by $L_\ell(k)$ in \eqref{eq:mismatch_loss}. In the local estimation regime, the corresponding RMSE inflation is approximately $1/\sqrt{L_\ell(k)}$. The predicted ratios of $1.004$, $1.019$, $1.045$, $1.124$, and $1.531$ are close to the reported ratios of $1.004$, $1.019$, $1.045$, $1.124$, and $1.527$. This agreement is consistent with correlation loss accounting for the observed reference-dependent difference, which is largest at $26.9$~m. The approximation need not hold in the threshold region or when peak selection varies across PRIs. Unlike physical-range reconstruction, velocity estimation contains no additive ambiguity-order term. Under the assumed constant-velocity model, range folding preserves the inter-pulse Doppler progression, with a constant phase absorbed into the unknown echo amplitude. This explains why Fig.~\ref{fig:rmse}(e) does not exhibit the metre-scale ambiguity-error behaviour of Fig.~\ref{fig:rmse}(b). Nevertheless, incorrect reference selection and delay errors can affect the amplitude and phase of the extracted slow-time sequence. At low SCNR, the departure from the local bounds is consistent with Doppler-peak selection errors. The results apply to point targets with constant radial velocity.

\subsection{Probability of Detection}
\label{subsec:detection_results}
Fig.~\ref{fig:rmse}(f) compares detection probabilities for the transmit and propagation-matched references at a per-cell false-alarm probability of $P_{\mathrm{FA}}=10^{-7}$. Under the fixed-cell Gaussian interference model of \eqref{eq:pd} and
\eqref{eq:pd_matched}, each reference uses a threshold scaled to its output interference variance. At the correct delay and Doppler, transmit-reference mismatch reduces the noncentrality through $\mathrm{SCNR}_\ell(k)=L_\ell(k)\mathrm{SNR}_\ell$.
The resulting horizontal separation on the common SCNR axis is $-10\log_{10}L_\ell(k)$, giving $0.04$, $0.16$, $0.38$, $1.02$, and $3.70$~dB for the five targets. The largest separation occurs at $26.9$~m, consistent with its larger correlation
loss. Initial detection in the proposed receiver uses the transmit reference. The matched-reference curves assume that the correct propagation reference is already available and therefore provide a detection benchmark. They do not establish the performance of an additional detection pass in the implemented receiver. Furthermore, the prescribed false-alarm probability applies to a fixed cell; the false-alarm probability over a searched delay--Doppler map also depends on the search size and cell dependence. The detection curves approach unity below approximately $-10$~dB, whereas all targets attain an empirical
identification probability of one only at $0$~dB and above. Detection is therefore less demanding than
ambiguity-order identification in the considered scene. Detection tests for the presence of an echo, while
identification must distinguish correlated candidate propagation responses. Their similarity contributes
to the higher identification requirement discussed in Section~\ref{subsec:perr}; however, \eqref{eq:pairwise} does not establish a universal SCNR penalty between the two tasks.

\section{Conclusion}\label{sec:conclusion}
This paper developed a receiver-side framework for resolving range ambiguity in high-PRF THz pulsed radar using
distance-dependent temporal broadening. Normalised correlations with candidate propagation-matched references are combined coherently across the CPI with Doppler compensation to identify the ambiguity order and reconstruct the physical range. The selected reference is reused for delay and Doppler estimation, reducing transmit-reference mismatch. Under known atmospheric conditions and with targets occupying distinct apparent-delay bins, simulations demonstrated range recovery beyond the conventional unambiguous interval. Range estimation errors conditioned on correct ambiguity-order
identification, together with the reported velocity estimation
errors, approached their corresponding Cramér--Rao bounds. Ambiguity-order identification governed the transition to accurate physical-range reconstruction and required higher SCNR than detection and local estimation. Compared with the evaluated diversity-based benchmarks, the proposed receiver required higher identification SCNR while preserving a fixed transmitted pulse shape, PRF, and sampling rate. These results demonstrate the numerical feasibility of exploiting THz propagation signatures for range disambiguation, with performance determined by candidate-reference separation, effective SCNR, and propagation-model accuracy.

\bibliographystyle{IEEEtran}
\bibliography{ref}

\end{document}